%% file: unreal-repair.tex
\documentclass[dvipsnames,format=sigconf,anonymous=false,review=false]{acmart}

\usepackage{booktabs} % For formal tables
\AtBeginDocument{%
  \providecommand\BibTeX{{%
    \normalfont B\kern-0.5em{\scshape i\kern-0.25em b}\kern-0.8em\TeX}}}

\copyrightyear{2023}
\acmYear{2023}
\setcopyright{acmlicensed}\acmConference[GECCO '23]{Genetic and Evolutionary Computation Conference}{July 15--19, 2023}{Lisbon, Portugal}
\acmBooktitle{Genetic and Evolutionary Computation Conference (GECCO '23), July 15--19, 2023, Lisbon, Portugal}
\acmPrice{15.00}
\acmDOI{10.1145/3583131.3590454}
\acmISBN{979-8-4007-0119-1/23/07}

\usepackage{graphicx,subfigure}
\usepackage{subfigure}
\usepackage{latexsym}
\usepackage{amsmath}
\usepackage{amssymb}
\usepackage{enumerate}
\usepackage{alltt}
\usepackage{url}
\usepackage{xcolor,colortbl}
\usepackage{multirow}
\usepackage{flushend}
\usepackage{proof}
\usepackage{array,multirow}
\usepackage{enumitem}

\renewcommand{\U}{\mathcal{U}}
\newcommand{\R}{\mathcal{R}}
\newcommand{\W}{\mathcal{W}}
\newcommand{\X}{\bigcirc}        %Next operator
\newcommand{\F}{\Diamond}        %eventually operator
\renewcommand{\G}{\Box}            %"always in the future" operator
\newcommand{\False}{\textit{false}}     %symbol for false
\newcommand{\True}{\textit{true}}      %symbol for true
\newcommand{\Implies}{\rightarrow}  %symbol for implication
\newcommand{\OurTool}{\textsc{AuRUS}}

\definecolor{darkgreen}{rgb}{0,0.65,0}

\usepackage{algorithm}
\usepackage{algpseudocode}
\algnewcommand{\LineComment}[1]{\State \(\triangleright\) #1}
\algrenewcommand\algorithmicindent{1.0em}

\pdfoutput=1
\makeatletter
\let\@authorsaddresses\@empty
\makeatother

\begin{document}

\title{Automated Repair of Unrealisable LTL Specifications Guided by Model Counting}
%\author{\IEEEauthorblockN{Anonymous Authors}}
\author[M.~Brizzio]{Matías Brizzio}
    \affiliation{%
    \institution{IMDEA Software Institute}
    \institution{Universidad Politécnica de Madrid}
    \country{Spain}}

\author[M.~Cordy]{Maxime Cordy}
    \affiliation{%
    \institution{SnT, University of Luxembourg}
    \country{Luxembourg}}
\author[M.~Papadakis]{Mike Papadakis}
    \affiliation{%
    \institution{SnT, University of Luxembourg}
    \country{Luxembourg}}

\author[C.~Sánchez]{César Sánchez}
    \affiliation{%
    \institution{IMDEA Software Institute}
    \country{Spain}}
\author[N.~Aguirre]{Nazareno Aguirre}
    \affiliation{%
    \institution{Universidad Nacional de R\'{\i}o Cuarto}
    \institution{CONICET}
    \country{Argentina}}
\author[R.~Degiovanni]{Renzo Degiovanni}
    \affiliation{%
    \institution{SnT, University of Luxembourg}
    \country{Luxembourg}}
\renewcommand{\shortauthors}{Brizzio et al.}


% \ccsdesc[500]{Software and its engineering~Software verification and validation}


\input{abstract}

\maketitle 

\input{introduction}
\input{background}
\input{motivation}

\input{approach}
\input{approximation}
\input{validation}
\input{related-work}

\input{conclusion}

\section*{Acknowledgment}
This work is made possible by the support of the Luxembourg National Research Funds (FNR) through the CORE project grant C19/IS/13646587/RASoRS, the PRODIGY Project (TED2021-132464B-I00) funded by MCIN/AEI/10.13039/501100011033/ and the European Union NextGenerationEU/PRTR, as well as a research grant from Nomadic Labs and the Tezos Foundation.
\newpage
\bibliographystyle{plain}
\bibliography{Bibliography}
% \newpage
% \input{appendix.tex}
\end{document}

%% file: abstract.tex
%!TEX root=unreal-repair.tex
\begin{abstract}
The reactive synthesis problem consists of automatically producing correct-by-construction operational models of systems from high-level formal specifications of their behaviours. However, specifications are often unrealisable, meaning that no system can be synthesised from the specification. To deal with this problem, we present {\OurTool}, a \textit{search-based} approach to repair unrealisable Linear-Time Temporal Logic (LTL) specifications. {\OurTool} aims at generating solutions that are similar to the original specifications by using the notions of \emph{syntactic} and \emph{semantic similarities}. Intuitively, the syntactic similarity measures the text similarity between the specifications, while the semantic similarity measures the number of behaviours preserved/removed by the candidate repair. We propose a new heuristic based on \textit{model counting} to approximate semantic similarity. We empirically assess {\OurTool} on many unrealisable specifications taken from different benchmarks and show that it can successfully repair all of them. Also, compared to related techniques, {\OurTool} can produce many \textit{unique} solutions while showing more scalability.
\end{abstract}

%% file: introduction.tex
%!TEX root=unreal-repair.tex
\section{Introduction}
\label{sec:introduction}

%\emph{Reactive systems} \cite{MannaPnueli1992} are characterised by the constant interaction with the environment in which they are executed.Typically, state transition systems are a well established formalism for modelling reactive systems behaviour, being central to a number of important activities in software development, such as model checking \cite{Clarke2001}, property monitoring \cite{DBLP:journals/tosem/BauerLS11}, and model based testing \cite{DBLP:journals/stvr/FraserWA09}. In these contexts, temporal logics are widely considered the standard formalisms in which expected properties of systems are expressed \cite{MannaPnueli1995,Clarke2008}. 

%State transition systems are an established formalism for modelling software system behaviour, especially in the context of \emph{reactive systems} \cite{MannaPnueli1992}, i.e., systems whose intended behaviour is characterised by the constant interaction with the environment in which they are executed. Indeed, the use of state transition systems as behavioural models of software is central to a number of important activities in software development, such as model checking \cite{Clarke2001}, property monitoring \cite{DBLP:journals/tosem/BauerLS11}, and model based testing \cite{DBLP:journals/stvr/FraserWA09}. In these contexts, temporal logics are widely considered the standard formalisms in which expected properties of systems are expressed \cite{MannaPnueli1995,Clarke2008}. 
Reactive synthesis is the problem of automatically generating a correct-by-construction implementation for a reactive system, from a given specification of the expected behaviour~\cite{MannaWolper1984,EmersonClarke1982,PnueliRosner1989,AlurLaTorre2001,Asarin+1998,DBLP:journals/jcss/BloemJPPS12}. 
%\changing{Briefly, a reactive specification contains a description of what are the expectations of the environment, together with a description of what are the system's goals or guarantees. }
The specification comprises the list of the variables controlled by the environment and system (i.e., inputs and outputs), respectively, and a declarative description of the expected properties of the system, i.e., the \emph{goals}. 
Based on these, reactive synthesis produces a model for the system, usually referred to as \emph{controller}, which interacts with the environment, and guarantees the specified goals~\cite{PnueliRosner1989}. 
 
Temporal logic is widely considered the standard formalism for expressing the expected goals of reactive systems \cite{MannaPnueli1992}. 
Reactive specifications are usually expressed as \emph{assume-guarantee assertions}, i.e., $A \Rightarrow G$, where $A$ captures \emph{assumptions} on the behaviour of the environment, and $G$ expresses the goals the system must fulfill, provided that the assumptions are met. 
Since the specifications are the central element in the synthesis, their quality is crucial for a successful process. 
% Since these specifications are the driving element in the synthesis process, their quality is essential to its success.

Reactive specifications often contain imperfections that make them \emph{unrealisable}, i.e., no controller can be synthesised. % that satisfies them. 
Common issues leading to unrealisability are (1) inconsistencies between goals making them unsatisfiable (and therefore unrealisable); 
%, and under an environment that satisfies the assumptions, the specification becomes unrealisable; 
(2) inadequate assumptions and guarantees, which allow the environment to satisfy the assumptions and prevent the system from complying with the guarantees. 
Therefore, arriving at a consistent \emph{and} realisable specification is not straightforward and demands comprehensive elicitation activities to prevent these common issues. 

Previous attempts have made significant efforts to provide automated mechanisms to assist engineers in identifying and resolving sources of unrealisability in temporal logic specifications. 
Some concentrate on \emph{diagnosing} the cause of synthesis impossibility, e.g., by computing a core of assertions that make the specification unrealisable~\cite{Cimatti:2008,Schuppan2010,DBLP:journals/sttt/KonighoferHB13}.
Some generate counter-strategies that evidence how the environment prevents the controller from satisfying the goals~\cite{RamanKress-Gazit2013}. 
%Other studies investigate how the imprecisions in assumptions and guarantees lead to inconsistent specifications~\cite{vanLamsweerde+1998,vanLamsweerdeLetier2000}, as well as the evolution in environmental conditions that demands the adaptation of system goals~\cite{Alrajeh+2020}. 
Related to our work, many approaches to repair unrealisable specifications have been presented~\cite{Chatterjee+2008,Alur+2013,Maoz+2019,CavezzaAlrajeh2016,li2011mining}. 
These techniques attempt to repair the specifications \emph{just} by adding assumptions that are built from the information extracted from the generated counter-strategies, and the current specification. 
This is a significant limitation because they do not consider that unrealisability might be caused by failures in the current assumptions and guarantees. 
%These existing techniques have several limitations. 
These techniques also impose syntactical restrictions on how specifications are written, limiting their application to particular patterns (e.g., GR(1), %which is not compatible with the general LTL specifications, \changing{GR(1) is 
a subset of LTL). 

In this paper, we present {\OurTool}, a \emph{search-based} approach to repair unrealisable specifications, which applies to LTL~\cite{MannaPnueli1992} specifications and generates candidate repairs by changing both, assumptions and guarantees. %,  with objectives following the lines of the above-mentioned studies. 
{\OurTool} consists of a genetic algorithm (GA) that, given an unrealisable LTL specification, attempts to generate a \emph{realisable} variant of it, which is as close as possible to the original one.
The algorithm iteratively explores candidate repairs of the original specification, %(by changing assumptions and guarantees), 
seeking to find a realisable variant that is both syntactically and semantically similar to the original one. 
The syntactic similarity is measured by using the number of \emph{sub-formulas} that %are not shared by% 
belong to both the original and the mutated specification. 
On the other hand, semantic similarity is measured by calculating the number of behaviours from the original specification that were maintained in the candidate repair.  
%belong to one of the formulas but not the other (i.e., number of traces satisfying one of the specifications but not the other). 
These behaviours are computed by using \emph{model counting}. 
Since all existing model counting approaches for LTL do not scale well, we develop an alternative approach to approximate the LTL model counting problem which considerably improves scalability. 
We empirically assess {\OurTool} and show that it is effective at repairing unrealisable specifications %, taken 
%at repairing all 26 unrealisable specifications provided. These specifications were collected 
%from the literature and synthesis benchmarks, 
not handled by previous techniques, while also producing more \emph{unique} repairs not computed by related approaches.
% Additionally, our approach was able to produce solutions which were previously elicited, manually, by users. 
%The remainder of this paper is organised as follows. Section~\ref{sec:background} introduces preliminary concepts. We motivate our work in Section~\ref{sec:motivating-example}.  We present the genetic algorithm in Section~\ref{sec:approach}. Section~\ref{sec:ltl-model-counting} describes how we estimate the problem of model counting for LTL. In Section~\ref{sec:validation} we report our empirical evaluation. Finally, we discuss related work in Section~\ref{sec:related-work} and present our conclusions in Section~\ref{sec:conclusion}.

%% file: background.tex
%!TEX root=unreal-repair.tex
\section{Preliminaries}
\label{sec:background}
\subsection{Linear-Time Temporal Logic (LTL)}
\label{sec:ltl}
LTL is a logical formalism widely employed to specify reactive systems~\cite{MannaPnueli1992}. Given a set $AP$ of propositional variables, LTL formulas are inductively defined using the standard logical connectives, and the temporal operators $\X$ (next) and $\U$ (until), as follows: 
\emph{(i)} every $p \in AP$ is an LTL formula, and \emph{(ii)} if $\varphi$ and $\psi$ are LTL formulas, then so are $\neg \varphi$, $\varphi \vee \psi$, $\X \varphi$ and $\varphi \U \psi$. 
Other connectives and operators, such as $\land$, $\G$ (always), $\F$ (eventually), and $\W$ (weak-until), can be defined in terms of the basic ones. 
LTL formulas are interpreted over infinite traces of the form $\sigma = s_0\ s_1 \ldots$, where each $s_i$ is a propositional valuation on $2^{AP}$. Formulas with no temporal operator are evaluated in the first state of the trace. Formula $\X \varphi$ is true in $\sigma$ iff $\varphi$ is true in $\sigma[1..]$, i.e., the trace obtained by removing the first state from $\sigma$. Formula $\varphi \U \psi$ is true in $\sigma$ iff there exists a position $i$ in the trace, such that, $\psi$ is true in $\sigma[i..]$ and for every $0 \le k < i$, $\varphi$ is true in $\sigma[k..]$. 
An LTL formula $\varphi$ is \emph{satisfiable} (SAT) iff at least one trace satisfies $\varphi$. 
The \emph{model counting} problem consists of calculating the number of models satisfying $\varphi$. 
In the case of LTL, if a formula is unsatisfiable, the number of models is zero. Otherwise, it has an infinite number of models.
% However, when an LTL formula $\varphi$ is satisfiable, then it has an \emph{infinite} number of models. 
Therefore, LTL model-counting 
is restricted to \emph{bounded models}~\cite{Finkbeiner+2014}, i.e., it computes how many models up to $k$ states exist for $\varphi$ and a bound $k$. 
% To guide the search, {\OurTool} relies on bounded model counting to quantify the semantic impact of the syntactic changes applied to the candidate repairs, measured as the number of behaviours preserved and removed from the original specification.
{\OurTool} uses bounded model counting to guide the search by measuring the semantic impact of syntactic changes on candidate repairs, in terms of the number of preserved and removed behaviors from the original specification.% with this number.

\subsection{Reactive LTL Synthesis}
\label{sec:gr1-synthesis}
Reactive LTL synthesis is the problem of automatically constructing a reactive module that 
reacts to the environment with the objective of realizing a given LTL 
specification of the form assume-guarantee, $\varphi: A \rightarrow G$~\cite{PnueliRosner1989}, defined over a set of variables $\mathcal{V} = \mathcal{X} \cup \mathcal{Y}$, where $\mathcal{X}$ and $\mathcal{Y}$ are the variables controlled by the environment, and system, respectively. 
A strategy for $\varphi$ is a function 
$\sigma: (2^{\mathcal{X}})^{+} \rightarrow 2^{\mathcal{Y}}$
that maps finite sequences of subsets of $\mathcal{X}$ into subsets of $\mathcal{Y}$. 
For an infinite sequence $X = X_1,X_2,\ldots \in (2^{\mathcal{X}})^{\omega}$, 
the play induced by strategy $\sigma$ is the infinite sequence $\rho_{\sigma,X} = (X_1 \cup \sigma(X_1))(X_2 \cup \sigma(X_2)))\ldots)$. 
A play $\rho$ is \textit{winning}
if $\rho \models \varphi$. A strategy is winning when 
$\rho_{\sigma,X} \models \varphi$ for all $X \in (2^{\mathcal{X}})^{\omega}$.

Realisability is the problem of deciding whether a specification has a winning strategy, and synthesis is the problem of computing one. 
The Unrealisability of a specification means that no winning strategy exists for the system. This implies that the environment can always falsify the specification, no matter which strategy the system chooses.
% I.e., \changing{the environment has a strategy} 
% \changing{Particularly, unrealisability} can be caused for many different reasons. 
{\OurTool} performs syntactic changes in the specification to remove the flaw that makes it unrealisable. 
{\OurTool} delegates the realisability check to Strix \cite{Meyer+2018}, one of the most efficient synthesis tools presented at the annual synthesis competition~\cite{SYNTCOMP}. 
\subsection{Genetic Algorithms}
\label{sec:genetic-programming}

Genetic algorithms \cite{Goldberg1989, Koza1992, Michalewicz1996} are heuristic search algorithms, inspired by natural evolution. 
%They have a number of distinguishing characteristics, compared to more traditional search algorithms. 
Candidate solutions are called \emph{individuals} or \emph{chromosomes}, and are often represented as sequences of \emph{genes} (characteristics) that capture their features. 
Genetic algorithms maintain a \emph{population} of candidate solutions, rather than a single ``current'' candidate, as in traditional search. 
They are largely driven by random decisions, e.g., in the generation of the initial population, and how the new candidate solutions are generated from existing ones. 
To produce new individuals, it exploits information in the current population, combining their characteristics (called \emph{crossover}), or randomly altering the information in specific individuals (called \emph{mutation}). 
The effectiveness of this general search process is guided by a heuristic function, called \emph{fitness function}. Intuitively, this function measures how ``fit'' a particular individual is, i.e., how close it is to being a real solution to the search problem under consideration. 
This evolution process is usually performed until some termination criterion is met, e.g., a defined number of iterations (known as \emph{generations} of the population). %, or  a sufficiently ``fit'' solution has been found.   
%As we describe later on in this paper, 
{\OurTool} employs genetic algorithms to search for realisable repairs, close to the unrealisable specification given as input. Individuals in our case will represent LTL specifications, the genetic operators will produce new specifications from others, and the fitness function will attempt to evaluate how ``close'' a candidate repair is to be realisable, as well as how close % an LTL specification 
is to the original (unrealisable) one. 
%For further details on genetic algorithms, we refer the reader to ~\cite{Michalewicz1996}. 

%% file: motivation.tex
%!TEX root=unreal-repair.tex
\section{A Motivating Example}
\label{sec:motivating-example}

Let us present a running example to illustrate the main ideas behind {\OurTool}. 
Consider the problem of synchronising the access to a shared resource, via an arbiter~\cite{KressTorfah2019}. Two processes request access to the resource via signals $r_1$ and $r_2$, respectively. An extra signal $a$ indicates when the resource can be accessed. The arbiter indicates which process has been granted access by means of respective signals $g_1$ and $g_2$.  Signals $r1$, $r2$ and $a$ thus constitute the \textit{inputs},  while signals $g1$ and $g2$ are the \textit{outputs}. The following guarantees are elicited in \cite{KressTorfah2019} for this problem:
\begin{small}
\begin{align*}
\pmb{G_1}:\G (r_1 \rightarrow \F g_1) \hspace{.7em} \pmb{G_2}:\G (r_2 \rightarrow \F g_2) \hspace{.7em} \pmb{G_3}:\G (\lnot a \rightarrow (\lnot g_1 \land \lnot g_2))
\end{align*}
\end{small}
% \centerline{$\pmb{G_1}:\G (r_1 \rightarrow \F g_1) \hspace{.7em} \pmb{G_2}:\G (r_2 \rightarrow \F g_2) \hspace{.7em} \pmb{G_3}:\G (\lnot a \rightarrow (\lnot g_1 \land \lnot g_2))$}
% }
Intuitively, $G_1$ states that if the first process requests access to the resource, the arbiter will eventually grant it. 
Guarantee $G_2$ states the same but for the second process. 
%Similarly, guarantee $G_2$ states that if the second process requests access, the arbiter will eventually grant it. 
%While 
While guarantee $G_3$  states that if the resource cannot be accessed, no process is granted permission. %While these are the guarantees, 
No assumptions were identified for this specification. 
Hence, the specification $S = G_1 \land G_2 \land G_3$ is \emph{unrealisable}. 

A cause of unrealisability is that the environment is allowed to set the input signal $a$ to $\False$ continuously, preventing the arbiter from granting access to the resource (see $G_3$). Therefore, if any of the processes requests access to the resource in such a situation, no implementation would satisfy the guarantees $G_1$ and $G_2$. 
In \cite{KressTorfah2019}, authors propose different alternatives for ``fixing'' these issues. %unrealisable specification 
%have been manually proposed. 
% \changing{In particular,} 
One option is to add an environment assumption to ensure that the resource is allowed to be accessed infinitely often; indeed, by adding the assumption $A_1 = \G \F a$, the resulting specification becomes realisable. 
Another option explored in~\cite{KressTorfah2019} is to indicate that the arbiter will enforce mutual exclusion in accessing the resource.
% i.e., it cannot grant access to the resource to both processes at the same time. 
This is done by replacing $G_3$ by an alternative guarantee $G_3' = \G(\neg (g_1 \land g_2))$. The resulting specification $S'=G_1 \land G_2 \land G_3'$ is  also realisable.
The overall intuition we get from this example is that, in cases of unrealisability, 
% this problem 
it can sometimes be fixed by making \emph{small} changes to the original specification. 
But, even when the %devised 
repairs were generated automatically, a domain expert would need to review them, and decide which of them is an acceptable solution. 
For instance, the repairs introducing a new assumption would need to be analysed to check if such an assumption is reasonable to expect from the environment. 
Thus, to help the domain expert and make easier this validation activity, we would like to maintain as much as possible from the original specification in the generated repairs, and simply modify what is necessary to make it realisable. 
% Similarly, ``weakening'' guarantees may lead to specifications that excessively divert from system requirements; if these are acceptable is an issue to be analysed by a domain expert. 
% 
% 
In this paper, we propose {\OurTool}, a genetic algorithm that performs syntactic modifications to an unrealisable specification, with the aim of producing a set of realisable candidate repairs. The candidate repairs are searched for in the \textit{``vicinity''} of the original specification, in the sense that they aim at being slight syntactic and especially semantic, modifications of the original one. {\OurTool} has the challenge of dealing with a very large search space of LTL specifications that are obtained by performing syntactic changes to the original one. 
It also needs to objectively quantify the semantic impact of each change. %, in terms of a semantic distance from the original specification. 
{\OurTool} is guided by a multi-objective fitness function that attempts to minimise syntactic and semantic changes, while at the same time attempting to achieve realisability. 
Let us provide some intuition on how {\OurTool} works, and how it can produce some solutions presented in \cite{KressTorfah2019}. 
%Let us provide some intuition on how our approach works, and in particular on how it is able to produce, for the above illustrating example, some of the solutions presented in \cite{KressTorfah2019}. 

{\OurTool} starts by generating an initial population that represents samples of candidate solutions. These are generated by introducing new assumptions based on patterns commonly found in reactivity specifications~\cite{Dwyer+1999}. For example, assumptions stating that input events occur infinitely often ($\G \F r_1$, $\G \F r_2$, $\G \F a$), and that different input events cannot simultaneously occur ($\G \neg (r_1 \land r_2 \land a)$), will be considered. In our running example, the initial population already contains one of the realisable solutions %to the arbiter specification, 
proposed in \cite{KressTorfah2019}. 
Other candidate solutions can be obtained by the successive application of genetic operators to some selected specifications. 
%Other candidate solutions based on more complex specification modifications can be obtained by the successive application of the genetic operators, that we describe below. 
% 
%To evolve the population at each iteration, genetic operators are applied  to some selected specifications. 
%To evolve the population, some candidate specifications are selected at each iteration, to which the genetic operators are applied. 
% 
{\OurTool} implements the two most common genetic operators, \textit{crossover} and \textit{mutation}. 
%Crossover works as follows. 
Given two specifications $S_1$ and $S_2$, the crossover operator will produce a new specification $S_3$ by replacing a sub-formula of $S_1$, by a sub-formula of $S_2$. For instance, if both $S_1$ and $S_2$ are exactly the same specification $G_1 \land G_2 \land G_3$, the crossover operator can produce a new specification $S_3 = G_1 \land G_2' \land G_3$, where $G_2'= \G(r_2 \Implies \F g_1)$ is obtained by replacing sub-formula $g_2$ in $G_2$, by the sub-formula $g_1$ extracted from $G_1$ in $S_2$. On the other hand, the mutation operator will create a new specification by applying a syntactic mutation to some sub-formula of the specification. 
%We support common changes applied in refining reactive specifications, like weakening or strengthening a formula, replacing connectives and temporal operators, etc. 
For instance, mutating generated specification $S_3$, the algorithm can produce a new specification $S_4 = G_1' \land G_2' \land G_3$ in which $G_1'= \G (r_1 \rightarrow \X g_1)$ is obtained by changing the operator $\F$ by $\X$ in $G_1$. 
Going back to the arbiter solutions, to obtain the realisable version $G_1 \land G_2 \land G_3'$, two mutations to the original guarantee $G_3$ are necessary: first, a replacement of the sub-formula $a$ by $\False$ leading to $\G (\lnot g_1 \land \lnot g_2)$; and then the replacement of operator $\land$ by $\lor$, obtaining the formula $\G (\lnot g_1 \lor \lnot g_2)$ which is equivalent to $G_3'$. {\OurTool} may produce other realisable specifications that can also be considered as candidate repairs. 
The fitness function plays a \emph{crucial} role in guiding the search. 
%Notice in particular that our genetic operators do not guarantee that all produced candidate repairs are indeed GR(1) specifications. 
%We later on describe how we deal with this issue, through the fitness function. 
% 
Intuitively, the \emph{fitness function} is the oracle that is used to assess the quality of the candidate solutions, giving higher scores to ``better'' individuals, i.e., those closer to sought solutions. {\OurTool} implements a multi-objective fitness function that assesses three key properties of the candidate solutions: \emph{(1)} it checks whether the specification is realisable or not; \emph{(2)} it then computes the syntactic similarity with respect to the original specification; and finally \emph{(3)} it computes the semantic similarity with respect to the original specification (i.e., the number of behaviours preserved and removed by the formula modifications). We 
% will show in the experimental section, that each one of these factors is very important for the fitness function to guide the search towards adequate solutions, i.e., solutions that are very similar to the original.
show in Section~\ref{sec:validation} the importance of each fitness factor to guide the search toward adequate solutions, i.e., solutions that are very alike to the original.

%% file: approach.tex
%!TEX root=unreal-repair.tex
\section{{\OurTool} Approach}
\label{sec:approach}
{\OurTool} takes as input an unrealisable specification $S$, and by the successive application of genetic operations, it aims at producing a specification $S'$ that is realisable, and minimizes the syntactic and semantic changes with respect to $S$.  %as similar as possible to $S$. 
\subsection{Search Space and Initial Population} 
\label{initial-pop} 
% Let $\mathcal{V}=\mathcal{X} \cup \mathcal{Y}$ be the disjoint set of variables of the original specification $S$. 
% where $\mathcal{X}$  denotes the variables controlled by the environment, and $\mathcal{Y}$ the variables controlled by the system. 
% Each individual of our search space is basically an LTL specification $S: A \Rightarrow G$ over $\mathcal{V}$, organized in terms of assumptions and guarantees. For clarity, we write $S = (A, G)$. 
% New individuals are obtained by the application of the genetic operators that produce syntactic modifications to both the assumptions and guarantees, with the \emph{same} probability
Individuals in our search space are LTL specifications $S = (A, G)$ over $\mathcal{V}$, consisting of assumptions and guarantees. Genetic operators are used to produce new individuals with syntactic changes to both assumptions and guarantees, with equal probability.
{\OurTool} begins by creating assumptions based on patterns commonly found in reactive specifications~\cite{Dwyer+1999} to form the initial population.
These new assumptions are generated from the original specification $S$%=(A, G)$
, resulting in $S_0 = (A \cup {a_0}, G)$, where $a_0$ follows the patterns:
(1) $\G \F x_i$, (2)$\G \neg (x_0 \land \ldots \land x_n)$, and (3) $\G \F (x_0 \land \ldots \land x_n)$, for $x_i \in \mathcal{X}$.
These patterns respectively express that the input $x_i$ holds infinitely many times, that all input events cannot happen at the same time, and that all input events hold at the same time, infinitely many times.
%(1) for each input $x \in \mathcal{X}$, $a_0 = \G \F x$ states that $x$ holds infinitely many times;  
%(2) for $x_i \in \mathcal{X}$, formula $a_0 = \G \neg (x_0 \land \ldots \land x_n)$ expresses all the input events can never by valid at the same time;
%and (3) for $x_i \in \mathcal{X}$, $a_0 = \G \F (x_0 \land \ldots \land x_n)$ states that all the input events hold at the same time, infinitely many times. 
% 
% REWRITED THIS
% Notice that, we only use input variables in the created assumptions, to avoid generating trivial solutions. For instance, if we added assumptions like $\G y$, $y \in \mathcal{Y}$, then such a specification would trivially become realisable, since it is possible to synthesize a controller that falsifies the assumption by setting the controlled variable $y$ to false, thus satisfying the specification. 
We only include input variables in the assumptions to prevent trivial solutions. Adding assumptions like $\G y$, $y \in \mathcal{Y}$ would result in a specification that can easily be satisfied by setting $y$ to false, making it trivially realisable.
% {\OurTool} takes care of not introducing this kind of anomalous solution (See~\cite{DIppolito+2013}) in the initial population, and also performs some checks while the search evolves (e.g., it does not produce repairs with unsatisfiable assumptions that trivially make the specification realisable). 
{\OurTool} avoids anomalous solutions~\cite{DIppolito+2013} in the initial population and checks for unsatisfiable assumptions during the search to prevent trivially realisable repairs.
%Notice that engineers can post-process the generated realisable solutions, to remove those with some undesired property (e.g., non-well separation~\cite{MaozRingert2016b}).
%Despite that our genetic algorithm takes care of not introducing this kind of anomalous solutions~\cite{DIppolito+2013} in the initial population, it does not perform any checking while the population and candidate repairs evolve in the search. 
%%RENZO: A comment this part that is more related to the implementation
% Additionally, in order to produce a more diverse initial population, we create some new specifications by the application of the random mutation operator to each assumption and guarantee of $S$, introduced later on in this section. For each assumption $a \in A$, we generate $S_0 = (A_0, G)$, where $A_0= A - \{a\} \cup \{a_0\}$ and $a_0 = mutate(a)$.  
% Similarly, for each $g \in G$, we generate $S_0 = (A, G_0)$, where $G_0= G - \{g\} \cup \{g_0\}$ and $g_0 = mutate(g)$.  
% Notice that the mutations applied to guarantees are not restricted to input variables; some controlled variable might be introduced too.
% 
% 
% 
% 
% 

\subsection{Genetic Operators} 
%To evolve the population, 
{\OurTool} implements the two most common genetic operators such as, \textit{crossover} and \textit{mutation}, adapted to LTL. % composing the assumptions and guarantees. 
%handle LTL specifications. In particular, our genetic operators will manipulate the LTL assertions composing the assumptions and guarantees. 
%Before presenting the genetic operators, we introduce some notation.
% Let us  first introduce some notation. 
Let $SF(\varphi)$ be the list of sub-formulas of $\varphi$, e.g., $SF(\G \neg p) = [\G \neg p, \neg p, p]$. 
We denote by $\varphi[\phi \backslash \psi] $ the formula that is obtained by replacing occurrences of $\phi$ in $\varphi$, by $\psi$. 
For instance, $\G(\neg p)[\neg p \backslash r]$ returns $\G(r)$. %, but $\G(\neg p)[q \backslash \neg r]$ returns the same formula $\G(\neg p)$ because $q$ is not a sub-formula of $\G(\neg p)$. 
Given two formulas $\varphi$ and $\varphi'$, we define: %the following functions:
$\it{replaceSub}(\varphi, \varphi') = \varphi[\phi \backslash \psi]$ and $\it{combineSub}(\varphi, \varphi') = \varphi[\phi \backslash \phi \bullet \psi]$, s.t. $\phi \in SF(\varphi)$,  $\psi \in SF(\varphi')$, and $\bullet \in \lbrace \vee, \wedge, \U, \R, \W \rbrace$. % is a binary operator. 
Intuitively, $\it{replaceSub}(\varphi, \varphi')$ returns a new formula that consists of replacing a sub-formula of $\varphi$ with a sub-formula of $\varphi'$; 
while $\it{combineSub}(\varphi, \varphi')$ takes a sub-formula from $\varphi$ and combines it with another from $\varphi'$, using a binary operator. 
% % 
% % 
 
The crossover operator combines two LTL specifications, namely $S_1 = (A_1, G_1)$ and $S_2 = (A_2, G_2)$, to create a new specification $S_3 = (A_3, G_3)$. This is achieved by merging the assumptions and guarantees of the original specifications.

% Particularly, $A_3$ is generated as follows:
% $A_3 = \{a | a \in A_1 \text{ or } a \in A_2 \text{ or } a=replaceSub(a_1,a_2) \text{ or } a=combineSub(a_1,a_2)\}$, where $a_1\in A_1$ and $a_2 \in A_2$.
% \noindent
%(1) $A_3 = A_1$; \hspace*{1em} (2) $A_3 = A_2$; \hspace*{1em}(3) $A_3 \subseteq A_1 \cup A_2$; \\
%(4) $A_3 = \{a | a \in A_1 \cup A_2 \ \lor a=replaceSub(a_1,a_2)\ \lor $\\
% 
% \noindent
Particularly, every assumption in $A_3$ is taken from $A_1$ or $A_2$, or is generated by replacing ($replaceSub(a_1,a_2)$) or combining ($combineSub(a_1,a_2)$) sub-formulas from $A_1$ and $A_2$, where $a_i \in A_i$. 
% One of the possibilities is that $A_3$ takes exactly the same set of assumptions from one of the specifications involved in the operation 
% (i.e., $A_1$ or $A_2$). 
% Another option is to merge $A_1$ and $A_2$, without necessarily taking all assumptions. 
% Cases in which \textit{replaceSub} and \textit{combineSub} are executed are more likely to produce new formulas appearing in $A_3$. 
% 
To produce $G_3$, the crossover operator performs the same choices but it takes guarantees from $G_1$ and $G_2$ instead.
% as for the assumptions but takes the guarantees from $G_1$ and $G_2$ instead. % of the assumptions. 
% 

The mutation operator takes a specification and applies syntactic modifications to produce a new one. 
Given a specification $S=(A, G)$, the algorithm randomly selects one assumption/guarantee to which the mutation will be applied. 
Let $mutate(\varphi)=\varphi'$ be the function that takes a formula and produces a mutation of it. 
When the mutation operator is applied to some assumption $a \in A$, it returns a new specification $S'=(A',G)$, where $A'= (A \setminus \{a\}) \cup \{a'\}$ and $a' = mutate(a)$. 
The same applies when some guarantee $g \in G$ is mutated. 
% the resulting specification will be $S'=(A,G')$, where $G'= (G  \setminus \{g\}) \cup \{g'\}$ and $g' = mutate(g)$. 
% The function $mutate$ applies random syntactical modifications to the formula. 
$mutate$ is defined as follows: %on the structure of $\phi$,
{%\fontsizeformula
\begin{enumerate}[leftmargin=*]
\item if $\phi = b$ or $\phi = p$, where $b \in \{\True,\False\}$ and $p \in AP$, then:
\begin{enumerate}
	\item $\phi' = b'$, s.t. $b' \in \{\True,\False\}$ and $b \neq b'$.
	\item $\phi' = q$, s.t. $q \in AP$ and $p \neq q$.
	\item $\phi' = o_{1} \phi$ where $o_{1} \in \lbrace \G, \F, \X, \neg \rbrace$.
\end{enumerate}
\item if $\phi = o_{1}\phi_1$, where $o_{1} \in \lbrace \neg, \X, \F, \G \rbrace$, then:
\begin{enumerate}
	\item $\phi' = mutate(\phi_1)$.
	\item $\phi' = o_{1}' mutate(\phi_1)$, s.t. $o_{1}' \in \lbrace \neg, \X, \F, \G \rbrace$. % and $o_1 \neq o_1'$.
	\item $\phi' = o_{1}' o_{1} mutate(\phi_1)$ where $o_{1}' \in \lbrace \neg, \X, \F, \G \rbrace$.
	\item $\phi' = p\ o_{2}'\ o_{1}'(mutate(\phi_1))$, s.t. $p \in AP$, \\ \hspace*{.6cm} $o_{2}' \in \lbrace \U, \W, \land, \lor \rbrace$ and $o_{1}' \in \lbrace \neg, \X, \F, \G \rbrace$.
	%\item $\phi' = p\ o_{2}'\ mutate(\phi_1)$, s.t. $p \in AP$ and $o_{2}' \in \lbrace \U, \W, \land, \lor \rbrace$.
	%\item $\phi' = p\ o_{2}'\ o_{1}(mutate(\phi_1))$, s.t. $p \in AP$ and \\ $o_{2}' \in \lbrace \U, \W, \land, \lor \rbrace$.
\end{enumerate}
\item if $\phi = \phi_1 o_{2} \phi_2$, where $o_{2} \in \lbrace \vee, \wedge, \U, \R, \W\rbrace$, then:
\begin{enumerate}
	\item $\phi' = mutate(\psi_{i})$, s.t. $\psi_{i} \in \lbrace \phi_1, \phi_2 \rbrace$ 
	\item $\phi' = mutate(\phi_1)\ o_{2}'\ mutate(\phi_2)$, s.t. $o_{2}' \in \lbrace \vee, \wedge, \U, \R, \W\rbrace$.
	\item $\phi' = o_{1}' (mutate(\phi_1)\ o_{2}'\ mutate(\phi_2))$, \\ \hspace*{.3cm} s.t. $o_{1}' \in \lbrace \neg, \X, \F, \G \rbrace$ and  $o_{2}' \in \lbrace \vee, \wedge, \U, \R, \W\rbrace$.
\end{enumerate}
\end{enumerate}
}

\subsection{Fitness Function}
\label{sec:fitness-fun}

{\OurTool} is guided by a multi-objective fitness function that aims at generating a realisable variant, as close as possible, to the unrealisable specification given as input. 
% To do that, the fitness 
This function focuses on checking three key properties of the candidate solutions, with the objective of finding one that is \emph{realisable}, and that minimises the \emph{syntactic} and \emph{semantic} changes with respect to the original specification.
%To do that, the fitness function focuses on checking the three key properties of the candidate solutions, i.e., whether the specification is realisable or not, and its semantic and syntactic similarity to the original unrealisable specification.
%with the objective of finding one that is \emph{realisable} and as similar as possible to the original unrealisable specification.
%and that minimises the \emph{syntactic} and \emph{semantic} distances with respect to the original unrealisable specification. 
Syntactic similarity is measured in terms of the number of sub-formulas shared by the original specification and the candidate. 
Semantic similarity is measured in terms of the number of behaviours that were maintained in the candidate. 
%that belong to one of the formulas but not the other. 
Basically, let $S$ be the unrealisable specification given as input, the fitness value for a candidate repair $S'$ is computed by the following function $f$: 

% {
% \begin{align*}
% f(S') = \alpha * status(S') + \beta * synSim(S, S') + \gamma * semSim(S, S')
% \end{align*}}
\noindent\hspace*{.25em}
\textit{\centerline{$f(S') = \alpha * status(S') + \beta * synSim(S, S') + \gamma * semSim(S, S')$}}

\noindent
where $status(S')$ focuses on checking if $S'$ is satisfiable and realisable; while, $synSim(S, S')$ and $semSim(S, S')$ compute the syntactic and semantic similarities between $S$ and $S'$, respectively. 
Constants $\alpha$, $\beta$ and $\gamma$ are the factors that assign different weights to the three properties of interest in the candidate repair $S'$. 

Let $S'=(A',G')$ be a candidate solution, we define $status(S')$ as follows:
\begin{small}
\vspace*{-1em}
\[
  \textit{status}(S') =
  \begin{cases}
        1 & \text{if $A' \land G'$ is satisfiable and $A' \rightarrow G'$ realisable;} \\
        0.5 & \text{if $A' \land G'$ is satisfiable, but $A' \rightarrow G'$ unrealisable;}\\
        0.2 & \text{if $A' \land G'$ is unsatisfiable, \text{but $A'/G'$ are satisfiable;}}\\
        0.1 & \text{if $A'$ is satisfiable, but not $G'$;} \\
        0 & \text{if $A'$ is unsatisfiable.} \\
  \end{cases}
\]
\end{small}
% Notice that, 
%$status(S')$ focuses on checking, satisfiability and realisability of $S'$. 
%According to ~\cite{Maoz+2019}, candidates that are realisable but unsat, are not \textit{useful repairs}.
$status(S')$ will return $1$ iff $S'$ is both, satisfiable and realisable. %, meaning that we have found an \emph{useful repair.} %for the originally unrealisable specification $S$.
When the candidate $S'$ is satisfiable, but still unrealisable, $status(S')$ will return $0.5$. 
% Also, $status(S')$ will return $0.2$ when the assumptions and guarantees are contradictory, $0.1$ if guarantees are unsatisfiable, 
Whether assumptions/guarantees are unsatisfiable, it will return values closer to $0$. 

$synSim(S, S')$ computes the \emph{syntactic similarity} between specifications $S$ and $S'$, measured in terms of the number of sub-formulas that belong to both the original $S$ and the candidate $S'$:
%Recall that $SF(\varphi)$ denotes the set of sub-formulas of $\varphi$, we define $synSim(S, S')$ as follows:
{%\fontsize{7.5}{7.5}
\begin{align*}
% synDist(S, S') = 0.5 * \#(SF(S) \cap SF(S')) \div \#(SF(S) \\
% + 0.5 * \#(SF(S) \cap SF(S')) \div \#SF(S')
synSim(S, S') = 0.5 * \left(\dfrac{\#(SF(S) \cap SF(S'))}{\#SF(S)}  + \dfrac{\#(SF(S) \cap SF(S'))}{\#SF(S')}\right)
\end{align*}
}

Small values for $synSim(S, S')$ indicate that $S'$ is syntactically very different from $S$, while values closer to $1$ indicate that both specifications are very similar. 
The fitness function uses this value to quantify the syntactic changes produced by the genetic operators. 

% However, even for small syntactic changes, the semantic impact can still be enormous (e.g., if the mutation operator negated the formula, from the syntactic point of view, there is only one syntactic change, but semantically the mutated formula characterises exactly the opposite behaviours of the original one). 
Small changes in the syntax of a specification may result in significant changes in its semantics. For example, a mutation that negates a formula may appear as a single syntactic change, but it can completely reverse the behaviors described by the original formula.
The function $semSim(S, S')$ computes the \emph{semantic similarity} between specifications $S$ and $S'$, measured in terms of the number of behaviours from the original specification still in the candidate repair. To automatically compute this value, we rely on LTL \emph{model counting}. Given a specification $S$ and a bound $k$, let us denote by $\#(S,k)$ the number of models up to $k$ states satisfying $S$. We define $semSim(S, S')$ as follows:\\
% \begin{align*}
\centerline{$semSim(S, S') = 0.5 * \left({\dfrac{\#(S\land S', k)}{\#(S, k)} + \dfrac{\#(S \land S', k)}{\#(S', k)}}\right)$}
% \end{align*}

Notice that small values for $semSim(S, S')$ indicate that the behaviours described by $S$ are very different from the ones described by $S'$. In particular, when $S \land S'$ is unsatisfiable, $semSim(S, S')$ is $0$. % (notice that we previously compute $status(S')$, so we know in advance if it is necessary, or not, to perform the model counting computations). 
As this value gets closer to $1$, both specifications characterise an increasingly large number of common behaviours. 
Since the computation of $semSim(S, S')$ requires calculating several model counting instances, it is crucial for {\OurTool} to perform model counting efficiently. Thus, later in Section~\ref{sec:ltl-model-counting}, we develop an automata-based estimation for LTL model counting, that scales much better than exact LTL model counting algorithms. 
The user can select different values for the different parameters that might affect the fitness function, e.g., $\alpha$, $\beta$, $\gamma$, and the bound $k$. 
% In Section~\ref{sec:validation}, we empirically assessed {\OurTool} with different configurations, in specifications of varying complexities, to select the best-performing combination. We also analyse its effectiveness in relation to previous techniques. 
We tested {\OurTool} in various complex specifications, selecting the best-performing configuration and comparing its effectiveness to previous techniques in Section~\ref{sec:validation}.
%, namely, $\alpha = 0.7$, $\beta = 0.1$, $\gamma = 0.2$, and $k = 20$, by default. 
%We selected these weights in such a way that $\alpha + \beta + \gamma = 1$. This means that the fitness value for each candidate solution is between $0$ and $1$, where better solutions are expected to have a value closer to $1$. 

\subsection{Selection}
% At each iteration, some individuals are selected to survive to the next generation in the evolution. 
% We use the traditional \emph{``best selector''} operator that consists of sorting the individuals of the current population by their fitness values in decreasing order, and selecting the best individuals until the maximum population size is reached. 
The fittest individuals are chosen for the next generation using the traditional \emph{``best selector''} operator, sorting individuals by fitness and selecting the best until maximum population size.

%{\color{red}{
%We assess the performance of our algorithm in Section~\ref{sec:validation}, in case studies of varying complexities, as well as its effectiveness in relation to other techniques. We also analyse the sensitivity with respect to the different configurations for the components of the fitness function.  
%}}

\subsection{Soundness, and (In)completeness}
%Let us now discuss the correctness and (in)completeness of our approach. 
% If {\OurTool} finds a solution, then it is guaranteed that the generated repair is both satisfiable and realisable. 
% %Particularly, the fitness function performs a number of checkings in order to determine if the candidate 
% {\OurTool} checks satisfiability using the LTL solver Polsat~\cite{DBLP:journals/corr/LiP0YVH13}, and realizability using Strix~\cite{Meyer+2018}. 
% %Thereby, {\OurTool} is guaranteed to produce only useful repairs~\cite{Maoz+2019}.
% %Then, by relying on the correctness of the LTL SAT solver used for this task, Aalta~\cite{Vardi+2015} in our case, and the synthesis tool used for checking for realisability, Strix~\cite{Meyer+2018}, our genetic algorithm is correct. 
% %It is also guaranteed to produce only \emph{useful repairs}~\cite{Maoz+2019}, i.e., satisfiable and realisable solutions. 

% {\OurTool} implements a non-exhaustive search and hence results to be incomplete, i.e., there may exist repairs that are not visited/considered. 
% However, our genetic operators are complete, i.e., given two formulas $\varphi$ and $\varphi'$ over the same vocabulary,  $\varphi'$ can be produced from $\varphi$ by the application of crossover and mutation. 
% Thus, all formulas over the same vocabulary can be produced by our algorithm.

{\OurTool} guarantees satisfiability and realisability of its generated repairs, checked using Polsat~\cite{DBLP:journals/corr/LiP0YVH13} and Strix~\cite{Meyer+2018}, respectively. While {\OurTool} uses a non-exhaustive search and may not consider all possible repairs, its genetic operators are complete, allowing the production of $\varphi'$ from $\varphi$ via crossover and mutation.

%% file: approximation.tex
\section{LTL Model Counting Approximation}
\label{sec:ltl-model-counting}
{\OurTool} proposes to use LTL model counting to  compute the semantic similarity $semSim(S, S')$. % between the initial unrealisable specification $S$ and a candidate repair $S'$, 
%Since it is part of the fitness function, it is evaluated frequently by the algorithm to guide the search, and thus we need to use an efficient mechanism % for solving LTL model counting. 
%As we explained in Section~\ref{sec:ltl}, model counting in the context of LTL is applied to bounded models. 
Unfortunately, exact LTL model counting techniques quickly reach their scalability limits~\cite{Finkbeiner+2014}. 
As a consequence, 
%some approaches tried to restrict their analyses either to safety-LTL %formulas 
%or to small bounded models. 
%However, none of those approaches is suitable for us, since we are dealing with general LTL formulas, including liveness, and on the other hand, we need model counting to scale to large bounds, as we want to compute the behaviors that $S'$ preserves and removes from $S$. 
the work presented in \cite{Degiovanni+2018} attempts to deal with the scalability issues. 
It presents a translation from LTL formulas to regular expressions, such that, each string recognized by the regular expression 
is a prefix of some trace satisfying the formula. 
Then, a string model counter (ABC~\cite{Aydin+2015}) is used to compute the number of strings that, up to a certain length, satisfy the regular expression. %(and that indirectly represents the number of prefixes that satisfy the LTL formula). 
Though effective, this approach works on specific examples but fails to scale overall 
(failing in 22/26 specifications, see Section~\ref{sec:validation}).
% (it failed in analysing 22 out of the 26 specifications, see  %considered in our experiments in 
 
%However, the process of generating the regular expressions, requires several expensive phases of minimization and determinization, that seriously affect its efficiency. We provide further details in the experimental evaluation in Section~\ref{sec:validation}. 
% that counting the number of \emph{bases}, corresponding to lasso traces satisfying a given LTL formula, is a good proxy to indirectly solve the LTL model counting problem. 

Thus, we develop a new technique to efficiently estimate the LTL model counting problem. 
Intuitively, we aim at counting the number of prefixes satisfying an LTL formula and use that number as a proxy for the number of models of the formula. 
To improve scalability, we employ matrices multiplication for this task (as ABC~\cite{Aydin+2015}). 

%Similar to bounded model checking \cite{Biere+1999}, and previous approaches on LTL model counting, here we restrict our analysis to a class of canonical models, called lasso traces. 
%Intuitively, a \emph{lasso trace} has the form $s_0 \ldots \ s_{i-1} (s_i \ldots s_k)^\omega$, where the states $s_0 \ldots s_k$ conform the \emph{base} of the trace, and the loop from state $s_k$ to state $s_{i}$ is the part of the trace that is repeated infinitely many times. 
%Then, instead of counting the number of lasso traces, up to a bound $k$, our idea is to count the number of bases of lasso traces of size $k$, and use this number as an estimation for the exact model counting number. 
%To do that, our approach starts by generating a finite state automaton $A_\varphi$, such that the set of paths recognised by $A_\varphi$ include the bases of all lasso traces satisfying $\varphi$. Intuitively, $A_\varphi$ recognises valid prefixes of traces satisfying $\varphi$. 
%Then, this automata will be encoded into a matrix $T_\varphi$, from which we can compute the number of prefixes of length $k$, that are accepted by the automaton $A_\varphi$. This value can then be used as an estimation for the LTL model counting problem.   \
% 
Basically, our approach operates in two steps.
First, we rely on established algorithms to generate an automaton $A_\varphi$ that recognises all the traces satisfying a given formula $\varphi$. 
Recall that LTL formulas are interpreted on infinite traces (c.f. Section~\ref{sec:ltl}).  
This means that every trace satisfying $\varphi$, has a path in automaton $A_\varphi$, in which some \emph{accepting state} is visited infinitely many times  (i.e., a loop). % in the path. 
Notice that, every \emph{finite} path reaching some accepting state of $A_\varphi$, is potentially a \emph{prefix} of some trace recognised by $A_\varphi$. 
Then, we can have an estimation of the number of traces accepted by $A_\varphi$, by counting the number of finite paths of $A_\varphi$ that reach some accepting state (this can be thought as if we interpreted accepting states as final states). 
Notice that, it may happen that our approach overestimates the number of models for $\varphi$. 
For instance, when some reachable accepting state does not loop. 
Conversely, the accepting state may be part of multiple loops; in this case, the prefix accepted by $A_\varphi$ corresponds to multiple traces, which leads to underestimating the number of models for $\varphi$. 
These are the reasons why our approach only \emph{approximates} the number of models.
%Precisely, our prototype uses the OWL library~\cite{Kretinsky+2018} to translate LTL to automata, and we actually use Parity automata, better suited for the synthesis context~\cite{Kretinsky+2018b}. 

Then, we encode automaton $A_\varphi$ into a $N \times N$ transfer matrix $T_\varphi$, where $N$ is the number of states in $A_\varphi$, such that the value of each $T_\varphi[i,j]$ denotes the number of transitions from states $i$ to $j$ in $A_\varphi$.   
The number of finite paths, of length $k$, reaching some accepting state of $A_\varphi$, can be computed by solving $I \times T_{\varphi}^{k} \times F$, where $I$ is a row vector codifying the initial states; $T_{\varphi}^{k}$ is the matrix resulting from multiplying $k$ times matrix $T_{\varphi}$; and $F$ is a column vector codifying the final (accepting) states of $A_\varphi$. 
% Second, our approach generates a Finite State Automaton $A_\varphi$ based on $B_\varphi$. 
% Basically, $A_\varphi$ is exactly the same as $B_\varphi$, with the main difference being in the acceptance condition. 
% While $B_\varphi$ recognises infinite traces,  $A_\varphi$ recognises finite traces, since its accepting condition requires reaching some final state. 
% 
% 
% Third, similarly to the model counting approach of ABC~\cite{Aydin+2015}, we encode $A_\varphi$ into a $N \times N$ transfer matrix $T_\varphi$, where $N$ is the number of states in $A_\varphi$, such that the value of each $T_\varphi[i,j]$ denotes the number of transitions from state $i$ to state $j$ in $A_\varphi$.   
% The number of traces of length $k$ accepted by $A_\varphi$ can be computed by solving $I \times T_{\varphi}^{k} \times F$, where $I$ is a row vector codifying the initial states; $T_{\varphi}^{k}$ is the matrix resulting from multiplying $k$ times matrix $T_{\varphi}$; and $F$ is a column vector codifying the final states of $A_\varphi$. 
%That is, $I[1,j] = 1$ if and only if $j$ is a initial state of $A_\varphi$, and $0$ otherwise; and $F[i,1] = 1$ if and only if $i$ is a final state $A_\varphi$, and $0$ otherwise. 
\begin{figure}
 \begin{minipage}[t]{0.22\textwidth}
 \centering
 \includegraphics[scale=.3]{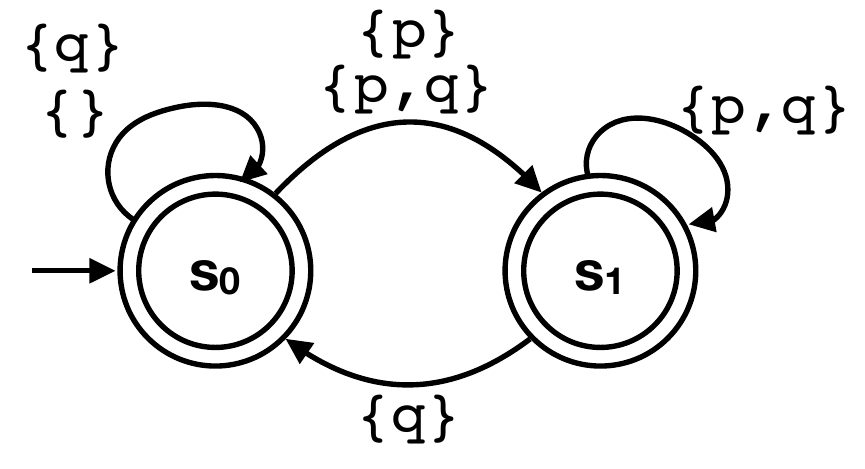}
 \caption{Finite automaton.}% for $\G(p \lor q)$.}
 \label{fig:dfa}
 \end{minipage}%\hspace*{-1em}
 \begin{minipage}[t]{0.23\textwidth}
 \centering
 \includegraphics[scale=.25]{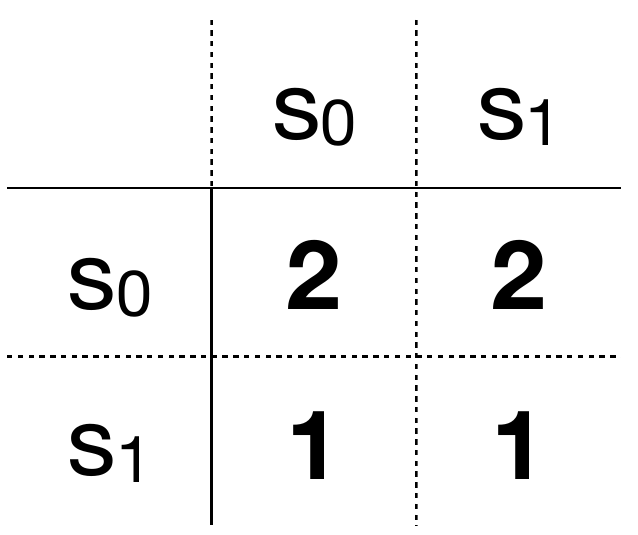}
 \caption{Transfer matrix.}% $\G(p \lor q)$.}
 \label{fig:transfer-matrix}
 \end{minipage}
 \end{figure}
For example, Fig.~\ref{fig:dfa} and Fig.~\ref{fig:transfer-matrix} show the automaton and the transfer matrix, generated by our approach from the formula $\psi = \G(p \Implies \X q)$. 
%For instance, considering the LTL formula $\G(p \Implies \X q)$ and 
Considering $k = 4$, our model counting  approach answers that there are 108 models, computed by the following matrices multiplication:
{
\[
I\times T_\varphi^4 \times F =
% \begin{bmatrix}
%     1  &  0  &  0  \\   
% \end{bmatrix}
% \times 
% \begin{bmatrix}
%     0  & 2  &  2  \\
%     0  & 1  &  1  \\
%     0  & 2  &  2  \\     
% \end{bmatrix}^4
% \times
% \begin{bmatrix}
% 	0	\\
%     1   \\
%     1      
% \end{bmatrix}
I
\times 
\begin{bmatrix}
    2  &  2  \\
    1  &  1  \\
\end{bmatrix}^4
\times
F
=
\begin{bmatrix}
    1  &  0   \\   
\end{bmatrix}
\times 
\begin{bmatrix}
    54  &  54  \\
    27  &  27  \\
\end{bmatrix}
\times
\begin{bmatrix}
    1   \\
    1        
\end{bmatrix}
= 108
\]
}
Actually, there are exactly 351 lasso traces of length 4 for $\psi$ % $\G(p \Implies \X q)$, 
but our approach reports 108 (i.e., the approximate number of prefixes for these traces). 
%{\color{red}{
However, we show in Section~\ref{sec:model-counting-results} that it provides a good estimation of the exact model counting. 
Meaning that, if it computes that formula $\varphi$ has more prefixes of length $k$ than formula $\psi$, then it is almost sure that the number of lasso traces of $\varphi$ of length $k$ is greater than the number of lasso traces for $\psi$. 

%% file: validation.tex
\section{Experimental Evaluation}
\label{sec:validation}
We evaluate {\OurTool} % based on the use of a genetic algorithm 
around the following research questions:
\begin{itemize}
\item[RQ1] \emph{How effective and efficient is {\OurTool}?} %in repairing unrealisable specifications?} 
\item[RQ2] \emph{How does it compare with random generation?}
% \item[RQ3] \emph{Does our approach generate unique solutions compared to related approaches?}
\item[RQ3] \emph{Does {\OurTool} generate unique solutions?}
\item[RQ4] \emph{How does each objective of the fitness function contribute to {\OurTool}' effectiveness?}
%\emph{How sensitive our approach, with respect to the three properties of interest of the fitness function, is?}
\item[RQ5] \emph{What is the precision of our model counting method?}
%\emph{Is our model counting method good at estimating the semantic distance between specifications?}
\end{itemize}
% To answer RQ1-RQ4, we consider several unrealisable specifications taken from the literature and existing benchmarks. %, that are accompanied by genuine realisable solutions. 
% Additionally, we use various randomly picked LTL formulas, taken from an LTL SAT solving benchmark, to answer RQ5.
We answer RQ1-RQ4 using unrealisable specifications from the literature and benchmarks, and RQ5 using randomly picked LTL formulas from an LTL SAT solving benchmark.

\textbf{Specifications.}
We consider 26 unrealisable specifications in our evaluation. 
%Table~\ref{tab:case-studies} summarises the size of each specification in terms of the number of input/output variables, assumptions (A) and guarantees (G). 
Table~\ref{tab:case-studies} summarises for each specification the number of input/output variables, assumptions (A), and guarantees (G). 
We consider 5 cases from the literature,
%: our motivating example Arbiter~\cite{KressTorfah2019}; a MinePump controller~\cite{Kramer+1983,Letier2001}; RG1 and RG2~\cite{Chatterjee+2008} that model mutual exclusion problems; and a Lift controller~\cite{Alur+2013}. 
% We also consider 
13 %unrealisable specifications 
from SYNTCOMP~\cite{SyntcompRepository}, %the synthesis competition (\changing{SYNTCOMP}) benchmark~\cite{SyntcompRepository}. 
and 8 specifications created by students and reported in SYNTECH15~\cite{Maoz+2019}. 
\begin{table}[htp!]
\caption{Unrealisable Specifications.}
\vspace*{-0.5cm}
\begin{center} 
%\scalebox{0.85}{
\resizebox{0.44\textwidth}{!}{
\begin{tabular}{|c|r|r|c|c|r|c|}
\cline{1-3}\cline{5-7}
Literature (5)	& \#In-Out &\#A-\#G & & SYNTCOMP (13) & \#In-Out &\#A-\#G \\
\cline{1-3} \cline{5-7}
Arbiter & 3-2 & 0-3		&& Detector & 2-1 & 0-4\\
MinePump & 2-1  & 1-2  	&& Full Arbiter& 3-3 & 0-16\\
RG1 	& 2-2	& 1-4 	&& Lily02 & 3-1 & 0-3 \\
RG2 	& 2-1	& 0-2 	&& Lily11 & 2-2 & 0-1\\
Lift 	& 3-3 	& 7-12 	&& Lily15 & 2-2 & 0-5\\
\cline{1-3} \multicolumn{3}{c}{}  && Lily16 & 3-3 & 0-9\\
\cline{1-3}
SYNTECH15 (8) & \#In-Out &\#A-\#G	&& Load Balancer & 3-2 & 3-8 \\
\cline{1-3}
Humanoid458& 3-10 & 0-11 && ltl2dba\_R\_2 & 2-1 & 0-1\\
Humanoid503& 6-11 &	1-17 && ltl2dba\_theta\_2 & 4-1 & 0-1\\
Humanoid531& 1-11 & 2-17 && ltl2dba27 & 1-1 & 0-1\\
Humanoid741& 4-14 & 5-21 && Prioritized Arbiter & 4-4 & 1-10 \\
Humanoid742& 1-14 & 2-26 && Round-Robin & 2-2 & 2-4\\
GyroV1 		& 3-3 & 6-7  && Simple Arbiter & 2-2 & 0-4\\	
\cline{5-7}
GyroV2 		& 3-3 & 7-7 	\\
PCarV2-888  & 3-9 & 4-21  	\\
\cline{1-3} 
\end{tabular}
}
\end{center}
\label{tab:case-studies}
\end{table}
\textbf{Implementation.}
{\OurTool} uses OWL~\cite{Kretinsky+2018} to manipulate LTL specifications. Apache Commons Math~\cite{CommonsMath} to manipulate matrices for model counting. 
{\OurTool} also integrates Polsat~\cite{DBLP:journals/corr/LiP0YVH13}, a portfolio that runs 4 LTL solvers in parallel.
%helping us to efficiently compute the fitness function. 
Moreover, {\OurTool} uses Strix~\cite{Meyer+2018} to check realisability.
The experiments in this section were conducted on a cluster with Xeon 2.6GHz, with 16Gb of RAM, running GNU/Linux. 
The tool, case studies, and a description of how to reproduce the experiments can be found in the replication package~\textbf{\url{https://sites.google.com/site/unrealrepair/}}. %The experiments were run on an Xeon 2.6GHz, with 16Gb of RAM, running GNU/Linux. 

\textbf{Experimental Setup.} 
As {\OurTool} is driven by random decisions, for each experiment, we run it 10 times. % on the selected case studies. 
%The tool supports different configurations for the genetic algorithm parameters, and some others related to the solver's configuration.
Precisely, in our experimentation {\OurTool} is configured as follows: the population size is 100, the model-counter bound $k$ is $20$, best selector, the crossover operator is applied to 10\% of the individuals, and the mutation operator is applied to each individual, to which each gene (sub-formula) is mutated with a probability of $1/N$ (where $N$ is the size of the formula). The termination criterion is reached either when 1000 individuals are generated or after 2hrs of execution time. %, whatever happens first.

Notice that, our fitness function (Sec.~\ref{sec:fitness-fun}) focuses on three aspects of the candidate solution (the status ($\alpha$), the syntactic ($\beta$) and semantic ($\gamma$) similarities). 
We assess the performance of {\OurTool} under many configurations, by considering values for $\alpha$, $\beta$, and $\gamma$, such that $\alpha + \beta + \gamma = 1$ (i.e., the sum of weights is 100\%). We organise our unrealisable specifications into two disjoint sets: 
the \emph{development} set, and the \emph{evaluation} set. 
We randomly selected 6 cases to be part of the development set  (2 from the literature, 2 from SYNTECH15, and 2 from SYNTCOMP) and the remaining 20 cases as part of the evaluation set. 
The development set is meant to support us in setting the parameters of our algorithm, with the hope that the best-performing configuration for the development set, will generalise to the evaluation set. 
%This will help us to assess how the results generalise from the development to the evaluation set, and to other case studies in the future.
To find the best-performing configuration, for each experiment we measure the number of repairs generated by {\OurTool}, as well as, the syntactic and semantic similarities of the found solutions. 
Then, for each case, we can compare the performance of two configurations by using the Vargha-Delaney A ($\hat{A}_{12}$) measure~\cite{VarghaDelaney2000}, to determine which configuration obtained better performance for that particular subject. % (in terms of number of solutions, syntactic and semantic similarities, respectively).
This will allow us to analyse which configuration generalises more and better to all the case studies. 
Particularly, realisability checking  is crucial for quality and quantity of solutions.
Also, better performance is reached when the weight assigned to the semantic similarity is greater or equal to the one assigned to the syntactic similarity. 
For instance, configurations $(\alpha=.7, \beta=.1, \gamma=.2)$, $(\alpha=.8, \beta=.07, \gamma=.13)$, or $(\alpha=.9, \beta=.05, \gamma=.05)$ typically reach better performance.
%$0.7 \leq\alpha$ and $\gamma \geq \beta$). 
%Then, the best configuration detected was: $\alpha=.7$, $\beta=.1$ and $\gamma = .2$.  
% 
% 
% 
%We run first 10 times our algorithm under the considered configurations, on the development set of specifications. 
%We selected the most effective configurations under the development set, and proceed to evaluate their effectiveness of the evaluation set.   
%From the set of all the runs, no matter the particular configuration, we collected the \emph{maximum} number of found solutions, that will be used as baseline to evaluate the relative performance of each configuration with respect to this ``ideal'' number of solutions.   
%Fig.~\ref{fig:fitness-configuration} summarises performance of our algorithm under several configurations. 
%Later on in  Section~\ref{sec:sensitivity-results} we study the sensitivity of our genetic approach with respect to the weights for $\alpha$ (realisability), $\beta$ (syntactic distance), $\gamma$ (semantic distance) that conform the fitness function, that helped us to find the best-performing configuration. 
\subsection{Effectiveness and Efficiency Evaluation}
\label{sec:experiments}
Table~\ref{tab:effectiveness} summarises the average results, out of the 10 runs, obtained with the best-performing configuration for each case study. % for our genetic algorithm. %, in which $\alpha=.7$, $\beta=.1$, and $\gamma=.2$. 
%Table~\ref{tab:effectiveness} summarizes the 10 runs.
% , where we denote by GA our approach.  %genetic algorithm. 
We report the average number of repairs and time (in seconds) per case required by {\OurTool} to explore the 1000 individuals.
% 
%Notice that the table includes all the cases studies from the development/evaluation sets.  
Noticeable, {\OurTool} succeeds in generating satisfiable and realisable repairs in 100\% of the runs. 
%Recall that our algorithm is guaranteed to produce only solutions that are satisfiable and realisable specifications, what~\cite{Maoz+2019} refers to as \emph{useful repairs}. 
As expected, it required more time to analyse the more complex specifications such as the Lift, full arbiter, and Humanoid cases in which the 2 hours timeout was reached. 
Particularly, there are four cases for which the algorithm could find just a few repairs: on average, 4 repairs for the full arbiter,  13 for the prioritized arbiter, and 3 repairs for Humanoid503 and PCarV2-888. 
This is because these cases contain several guarantees that require many changes by the algorithm to finally find realisable solutions. 
In particular, the first 2 cases are part of the synthesis competition and were made artificially unrealisable (by adding assertions contradicting the existing ones) to use them for assessing the efficiency of the tools participating in the competition. 
%. We did not performed any modification to the specifications, these extra constraints are part of the synthesis competition benchmark, to make the specifications unrealisable, to use them for assessing the efficiency of the solvers participating in the competition.
%Then, we validate the found solutions with respect to the genuine versions, previously identified. 
%Out of the 10 runs, we report the number of times that the genetic algorithm succeeded in finding at least one genuine solution (\#Suc.) and the average number of genuine solutions generated (whether more than one genuine solution has been identified). Similarly, we report the number of times that we succeeded in finding weaker (stronger) solutions and the average number of weaker (stronger) solutions found. 

%\vspace*{-0.25cm}
\begin{table}[htp!]
\caption{Comparison between {\OurTool} and random.}
\vspace*{-0.5cm}
\begin{center} 
%\scalebox{.87}{
\resizebox{.44\textwidth}{!}{
\begin{tabular}{|c|rrr|c|c|rrr|}
\cline{1-4}\cline{6-9}
%& \multicolumn{3}{c|}{Analysis} && \multicolumn{3}{c|}{Analysis}\\
Literature 	& Tech. & \#Sol. & Time && SYNTCOMP	& Tech. & \#Sol. & Time \\
\cline{1-4}\cline{6-9}
Arbiter & {\OurTool} 		& 467 	& 921 	&& Detector	& {\OurTool} 	 & 522 & 1799	\\ 
		& Random 	& 11 	& 404 	&&  		& Random & 21  & 1592\\
\cline{1-4}\cline{6-9}
Minepump & {\OurTool} 	 	& 481 	& 897	&& Full Arbiter & {\OurTool}  & 4 & 3805\\
		 & Random 	& 31  	& 678	&& 			& Random  & 6 & 1003\\
\cline{1-4}\cline{6-9}
RG1 	& {\OurTool} 		& 380 	& 905	&& Lily02 	& {\OurTool}  & 387 & 2656\\
 		& Random 	& 15	& 529 	&& 			& Random  & 4 & 427\\
\cline{1-4}\cline{6-9}
RG2 	& {\OurTool} 		& 459 	& 935	&& Lily11 	& {\OurTool}  & 623 & 834\\
 		& Random 	& 27 	& 406 	&& 			& Random  & 35 & 350\\
\cline{1-4}\cline{6-9}
Lift 	& {\OurTool} 		& 303 	& 3170 	&& Lily15 	& {\OurTool}  & 424 & 1643\\
	 	& Random 	& 0 	& 930 	&& 			& Random  & 4 & 1248\\
\cline{1-4}\cline{6-9}
\multicolumn{4}{c}{}  				&& Lily16 	& {\OurTool} & 385 & 1756 \\
\cline{1-4}SYNTECH15 	& Tech. & \#Sol. & Time && & Random & 6 & 984 \\
\cline{1-4}\cline{6-9}
GyroV1 	& {\OurTool} 		& 530 	& 1574 	&& Load Balancer 	& {\OurTool}  & 532 & 1619\\
		& Random 	& 7  	& 724  	&& 			& Random & 29 & 801 \\
\cline{1-4}\cline{6-9}
GyroV2 & {\OurTool} 		& 618 	& 1388 	&& ltl2dba\_R\_2 	& {\OurTool} & 623 & 1442 \\
		& Random 	& 40 	& 771 			&& 			& Random & 42 & 1451 \\
\cline{1-4}\cline{6-9}
Humanoid458 & {\OurTool} 	& 582	& 2307 	&& ltl2dba\_theta\_2 	& {\OurTool}  & 660 & 1453\\
		& Random 	& 26 	& 738	&& 			& Random & 32 & 1493 \\
\cline{1-4}\cline{6-9}
Humanoid503 & {\OurTool} 	&	3	&	7400	&& ltl2dba27 & {\OurTool} & 582 & 1473 \\
		& Random 	&   1	& 	616	&&	& Random & 46 & 1048 \\
\cline{1-4}\cline{6-9}
Humanoid531 & {\OurTool} 	&	86	&	7400	&& Prioritized Arbiter 	& {\OurTool}  & 13 & 4205\\
		& Random 	&  	19  & 	 239&& 			& Random & 24 & 5680 \\
\cline{1-4}\cline{6-9}
Humanoid741 & {\OurTool} 	&  80	&	7400	&& Round-Robin 	& {\OurTool}  & 678 & 1713\\
		& Random 	&  11	& 	756 && 				& Random  & 76 & 904 \\
\cline{1-4}\cline{6-9}
Humanoid742 & {\OurTool} 	&	99	&	7400	&& Simple Arbiter & {\OurTool}  & 504 & 1012\\
		& Random 	&  	21  & 	705 && 				& Random & 17 & 404 \\
\cline{1-4}\cline{6-9}
PCarV2-888	& {\OurTool} 	&	3	&	7400\\
		& Random 	&  	4  & 	174\\
\cline{1-4}
\end{tabular}
}
\end{center}
\label{tab:effectiveness}
\end{table}

\vspace*{-0.5cm}
\subsection{Comparison with Random Generation}
\label{sec:random-results}
Random starts by producing 1000 syntactic variants of each unrealisable specification and then checks which one is satisfiable and realisable. 
The random approach uses the same mutation operator as {\OurTool} to produce syntactic modifications to the original specifications, with the additional requirement that at least one sub-formula (assumption/guarantee) has been modified. 
We repeat this experiment 10 times and report all results in Table~\ref{tab:effectiveness}.
Notice that, unsurprisingly, the time required by random is considerably smaller than the required by {\OurTool} in most cases, except for cases ltl2dba\_R\_2, ltl2dba\_theta\_2 and Prioritized Arbiter.
However, random effectiveness is relatively low compared to {\OurTool}, producing on average 23 times less repairs than {\OurTool}.  
We also compute, for each case study, the $\hat{A}_{12}$ measure to compare the number of repairs obtained in the 10 runs of our best-performing configuration and random. 
$\hat{A}_{12}$ results to be 100\% for almost every case study, indicating that {\OurTool} obtains more repairs than random in every run. 
There are only 3 exceptions where random produces more realisable solutions: in Full Arbiter ($\hat{A}_{12}$ of 33.3\%), Prioritized Arbiter ($\hat{A}_{12}$ of 10.5\%) and PCarV2-888 ($\hat{A}_{12}$ of 33.3\%). 
Fig.~\ref{fig:comparion} shows further details regarding random performance.

%%% I COMMENT THIS BECAUSE WE DON NOT KNOW IF OUR APPROACH ALSO GENERATES UNDERISED SOLUTIONS
%However, when we manually analysed these solutions, we observed that the random added basic assumptions, using only output variables, to repair  these two cases in consideration. This allows to the controller to trivially satisfy the specification, by falsifying the assumptions. This is a typical problem in reactive system specification, called non-well separation~\cite{KleinPnueli2011,MaozRingert2016b}, in which the specification can be trivially realised by the controller. 
%This observation needs further analysis, and we plan as future work systematically studying how frequent random generation, and our genetic algorithm, produce non-well separated repairs. 

%Moreover, our algorithm produces candidate solutions that are closer to the original intentions of the specifier (i.e., equivalent, slightly weaker or stronger than the genuine ones), while at the same time providing many more choices from which to select how to refine an unrealisable specification. 

\subsection{Comparison with Related Approaches}
\label{sec:related-work-comparison}
%Here we study if there is or not some relation between the solutions produced by our approach and 
We study to what extent {\OurTool}'s repairs are unique or related to the repairs produced by other approaches.  
Precisely, we analyse if {\OurTool} is able to produce some equivalent, weaker, or stronger repair. 
We say that a formula $B$ is weaker than $A$, if $A \rightarrow B$ holds (i.e. if $A \land \neg B$ is unsatisfiable). 
% We can check this automatically by using satisfiability: if $A \land \neg B$ is unsatisfiable, then  $A \Rightarrow B$.  
Typically, weaker specifications are thought of as more general solutions, while stronger ones correspond to more localised solutions. 
Our intention is twofold: we want to show that {\OurTool} can produce repairs that are close to the ones obtained by other approaches, and also that it can generate \textit{unique} solutions that cannot be computed by existing techniques.  
%Weaker and stronger solutions are interesting and useful too, as they allow engineers to explore alternative solutions, until an appropriate specification is obtained. 

\emph{Manually reported solutions.} 
%We focus on manual solutions reported in the literature.  
{\OurTool} can generate \emph{equivalent} solutions to some manual repairs, reported in the literature, for the Arbiter, MinePump, RG2, and Lift cases. %, but no one for the RG1. 
Additionally, it produces some \emph{weaker}/\emph{stronger} solutions, compared to the manual ones, as well as many unique solutions, giving further choices to the engineer in how to refine the specifications to get a realisable one. 

%As expected, the algorithm requires more time to analyse the four more complex examples (i.e., Lift, GyroV1/2 and Humanoid), and despite that it succeeded in finding realisable repairs for each one of them, the number of solutions learnt that are related to genuine solutions, is relatively small. We remark that, when translating specifications from Spectra to TLSF, OWL changed the representation of some variables, e.g., enumerative types in Spectra are characterised as boolean combinations in TLSF. This might be affecting the performance of our algorithm. We plan to investigate if by improving this translation, we can improve the quality of the learnt repairs for these cases. 

\emph{Automatically generated solutions.} 
% For this comparison, 
We consider the work presented by Maoz et al.~\cite{Maoz+2019}, limited to GR(1). 
% of LTL specifications, called GR(1) -- General Reactivity (1). 
They present two symbolic techniques for learning missing assumptions: \emph{JVTS-Repair}, which generates new assumptions from the counter-strategies built as proofs of unrealisability; 
and \emph{GLASS}, that computes safety, justice, and initial assumptions to ensure the corresponding safety, justice, and initial guarantees of the unrealisable specification. 
Also, \cite{Maoz+2019} re-implements the algorithm presented in~\cite{Alur+2013}(\emph{AMT13}), which also generates missing assumptions from counter-strategies.
Notice that, while \emph{GLASS} generates only 1 candidate repair, \emph{JVTS-Repair} and \emph{AMT13} may generate many candidates (because they try to remove the counter-strategies generated). 
%From our dataset, only 1 specifications are expressed into GR(1) fragment: RG1, RG2, Lift, and the 8 cases in \changing{SYNTECH15}. 
To perform this comparison we took specifications from SYNTECH15 and others expressed in GR(1) such as RG1, RG2, and Lift (see Table~\ref{tab:case-studies}).

%Notice that in \cite{Maoz+2019}, more unrealisable GR(1) specifications were considered, but unfortunately these cannot be analysed by our prototype, due to it does not support past-time LTL operators (a common practise to encode LTL specifications into GR(1), if possible, is to use auxiliary variables and past-time constraints). 
%Thus, we restrict our comparison to the mentioned 6 case studies. 
%Due to the techniques of \cite{Maoz+2019} only produce changes in the assumptions, to make a fair comparison, in this case we do not consider the solutions found by our approach in which some guarantee is modified. 
%\emph{Comparison with GLASS:}
\begin{table}[tph!]
\caption{Repairs Overlapping}%: we report the number of unique (left) and equivalent (right) solutions produced by each technique.}
\vspace*{-0.5cm}
\begin{center} 
%\scalebox{.65}{
\resizebox{.34\textwidth}{!}{
\begin{tabular}{|c|r|r|r|r|c|}
\hline
Case	& {\OurTool} & GLASS & JVTS-Repair & AMT13\\
\hline
RG1 	& 379 / 1  & 1 / 0 & 46 / 0 & 98 / 1\\
RG2 	& 453 / 6  & 0 / 1 & 11 / 1 & 20 / 4\\
Lift 	& 302 / 1  & 1 / 0 & 145 / 0 & 1 / 1\\
GyroV1  & 529 / 1  & 1 / 0 & 7 / 0 	& 22 / 1\\
GyroV2 	& 617 / 1  & 1 / 0 & 9 / 0 	& 13 / 1 \\
Humanoid458& 582 / 0  & 1 / 0 & 1 / 0 	& 1 / 0 \\
Humanoid503& 3 / 0  & 1 / 0 & Timeout & Timeout\\
Humanoid531& 86 / 0 & 1 / 0 & Timeout & Timeout\\
Humanoid741& 80 / 0 & 1 / 0 & Timeout & Timeout\\
Humanoid742& 99 / 0 & 1 / 0 & 3 / 0 & Timeout\\
PCarV2-888 & 3 / 0 & 1 / 0 & 289 / 0 & Timeout\\
\hline
\end{tabular}
}
\end{center}
\label{tab:comparison-related-work}
\end{table}

Table~\ref{tab:comparison-related-work} summarises, for each case, the number of unique solutions (left) produced by {\OurTool}, i.e., the number of solutions not generated by other techniques, and the number of equivalent solutions to one produced by other approaches (right). Notice that the comparison is always in between {\OurTool} and the related techniques, but we do not compare the related approaches against each other (i.e., we do not compare \emph{GLASS} vs \emph{JVTS-Repair} vs \emph{AMT13}). 
% \changing{Table~\ref{tab:comparison-related-work} must be read as follows: For the {\OurTool} column, the number of equivalent and unique solutions is related to the other three techniques as a whole. On the other hand, the numbers shown in other columns are only related to {\OurTool}.}
%Table~\ref{tab:comparison-related-work} reads: For the {\OurTool} column, the number of equivalent and unique solutions is related to the other three techniques. On the other hand, the numbers shown in other columns are only related to {\OurTool}
%On the other hand, the numbers shown in the column of each technique such as GLASS, JVTS-Repair, and AMT13 are only related to GA.}
%Moreover, between parenthesis we indicate the number of non-unique solutions generated for our approach, we indicate between parenthesis the number of solutions that are equivalent to some solution generated by the other techniques. Similarly, for the other techniques, we indicate the number of solutions equivalent to some of ours. 

We observe that in most of the cases, {\OurTool} and \emph{GLASS} complement each other, with the only exception for case RG2, where {\OurTool} produces an equivalent solution to the one proposed by \emph{GLASS}. 
Moreover, for all considered cases, {\OurTool} generates many weaker/stronger repairs (between 2 and 80) than the one provided by GLASS, and the remaining are unique. 
%This implies that, the repairs generated by our approach that include stronger assumptions, result to be \emph{more general} solutions than the one provided by GLASS (notice that assumptions are part of the antecedent in the specification $A \Rightarrow G$, thus, the stronger the assumptions, the weaker the specification). 
%When we analysed JVTS-Repair and AMT13, the results are more varied, since these approaches generate several repairs.
% 
When analysing \emph{AMT13} repairs, we observe that {\OurTool} generates, for all cases in which the bound of \textit{10 minutes} was not reached (same timeout of~\cite{Maoz+2019}) except for Humanoid458, between 1 to 4 equivalent solutions, and the remaining  
In the case of \emph{JVTS-Repair}, {\OurTool} only generates 1 equivalent solution for the case RG2.
In all case studies, % between 5 and 100 of the 
many solutions generated by {\OurTool} maintains some relation (i.e., weaker/stronger) to the ones generated by JVTS-Repair and AMT13, but the majority are unique. 
The results evidence that the overlapping between the solutions is low, indicating that {\OurTool} can 
%be used to 
\textit{complement} existing techniques, and provide a rich set of variants to the engineer to resolve the source of unrealisability. 
%The previous observations suggest that our approach can be used to complement the existing techniques, and provide more variants to the engineer to resolve the source of unrealisability. 

%One of the limitation of our approach, is that currently it does not support past-time LTL operators. 
%On the other hand, our approach applies to general LTL reactive specifications, without imposing any syntactical restriction on the specifications, and aims at solving the cause of unrealisability by performing changes in both, the assumptions and guarantees.
%Some of them were manually produced, while others were automatically generated using analysis tools. It is worth mentioning that we do not perform any change to these specifications, neither to the unrealisable nor to the genuine realisable version. We have however to translate some of the specifications from Spectra~\cite{SpectraRepo} to TLSF, the input language of our prototype, for which task we use OWL~\cite{Kretinsky+2018}. 

\subsection{Importance of the three properties} %in the fitness function}
\begin{figure}[tph]
\centering
\includegraphics[scale=.28]{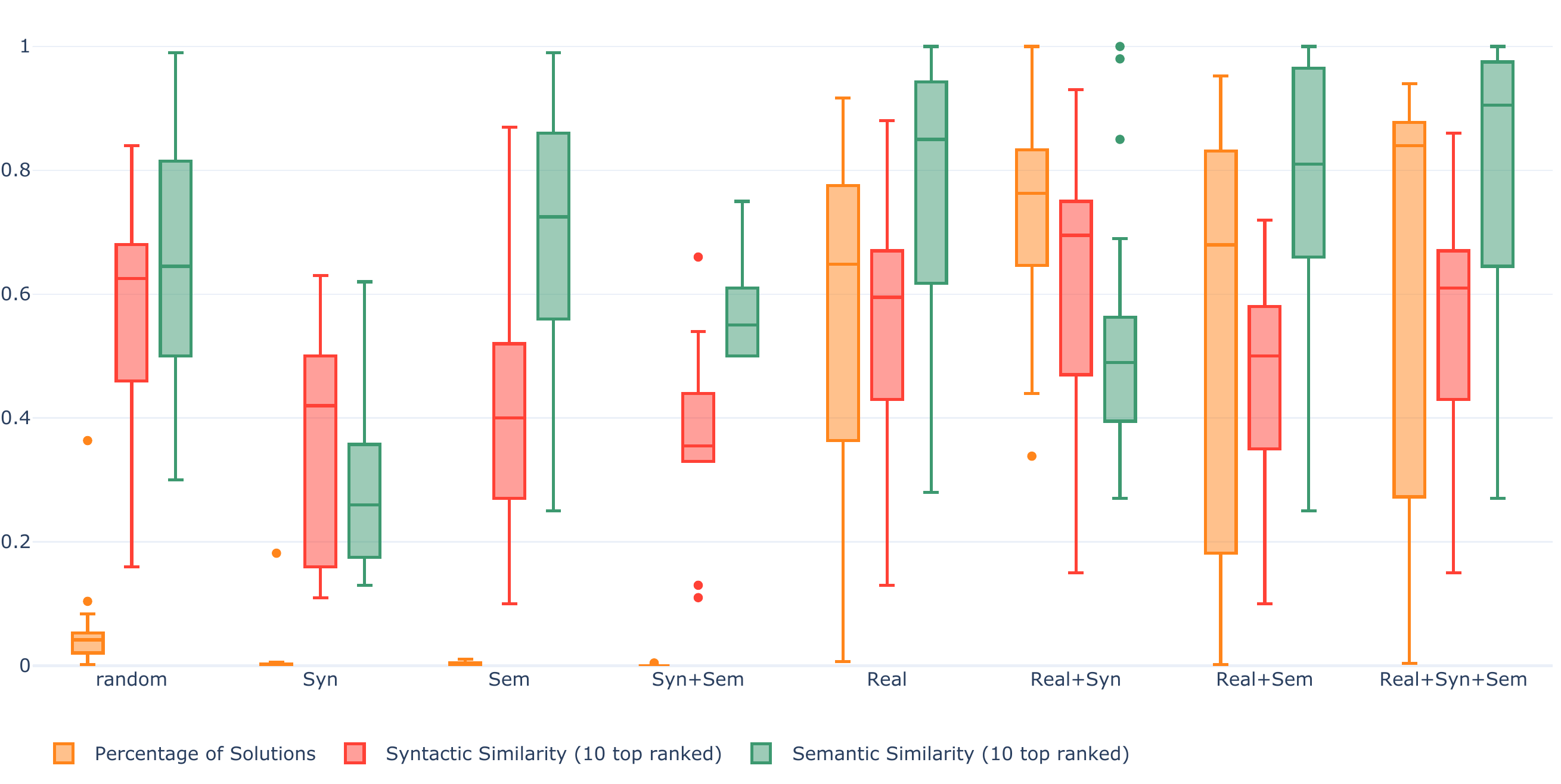}
\caption{Impact of each factor of the fitness.}% in the effectiveness of our approach.s}
\label{fig:comparion}
\end{figure}

\label{sec:sensitivity-results} 
%In this section we asses the importance of the three aspects on which our fitness function focuses, namely, the realisability status (Real), the syntactic (Syn) and semantic (Sem) distances of candidate repairs to the original unrealisable specification. 
We first study {\OurTool}' effectiveness, when some properties are deactivated from the fitness function.  
We run it under six extra configurations (see Fig.~\ref{fig:comparion}). 
In configurations (Syn, Sem, Syn+Sem) we deactivate the status checking, being {\OurTool} only guided by the syntactic and/or semantic similarity (realisability is only checked at the end of the execution, to check which candidate is a solution).  
In configurations (Real, Real+Syn, Real+Sem) we deactivate the syntactic and/or semantic similarity computation from the fitness function. Configuration Real+Syn+Sem denotes that {\OurTool} is guided by the three properties. 
% 
%For each case study, we run the algorithm 10 times with each one of these configurations. 
Fig.~\ref{fig:comparion} reports, for each configuration, the average percentage of repairs found per case study, with respect to the best performance previously discussed (with the three factors activated). 
Precisely, for the orange plots (repairs produced), the y-axis represents the relative difference between configuration runs and the best result for this metric across all configuration runs. For instance, the configuration \textit{Real+Syn+Sem} produced on average 467 repairs while the highest number across all configurations is 531 in the arbiter example. 
In the red and green plots, the similarity is measured relative to the original specification.
Notice that, by removing the realisability checking, {\OurTool} behaves pretty similar, or even worse, than random, affecting drastically its effectiveness in finding repairs. 
On the other hand, configurations that use realisability checking considerably improve {\OurTool}' effectiveness, being able to find more solutions. 
In fact, we compute $\hat{A}_{12}$ values to compare different configurations for each case, and we can ensure that if {\OurTool} is guided by the 3 properties (\textit{Real+Syn+Sem}), it obtains more repairs than the configurations that remove some of the properties. 

We also analyse the quality of the best 10 ranked repairs, 
measured in terms of the syntactic (red) and semantic (green) similarity,   %, reported in Fig~\ref{fig:comparion}.  
that can be presented to the engineer for analysis and validation. 
It is almost always the case that our best-performing configuration (Real+Syn+Sem) obtains better syntactic and semantic similarities than the other configurations. 
Only in a few cases where random found more repairs than {\OurTool}, 
outperforms us in syntactical and semantic similarities. 
Two exceptions occur in Humanoid741/742, where the solutions found by configuration Real+Sem have better semantic similarities than the repairs found by Real+Syn+Sem. 
%, allowing the user to count with more choices to refine the specification. 
The outliers in Fig.~\ref{fig:comparion} correspond to the mentioned cases. % in which {\OurTool} was not very effective.

%, full arbiter and prioritized arbiter, in which our algorithm was not very effective.
%whose quality have been affected as well, and consequently, the number of genuine versions found is considerable reduced.

%\vspace{-1em}
%\vspace{-2em}
% \begin{figure}[htp]
% \centering
% \includegraphics[scale=.47]{plots/configuration.pdf}
% \caption{Performance under different configurations.}% for $\G(p \lor q)$.}
% \label{fig:fitness-configuration}
% \end{figure}

\subsection{Evaluating our Model Counting Approach}
\label{sec:model-counting-results}
To evaluate the precision and scalability of our model counting approach, we compare it with two related approaches. % for the comparison. 
Firstly, we re-implemented an established exact model counting approach, that will be used as a baseline in the comparison. 
We basically encode LTL formulas as propositional formulas, such that, for a given bound $k$, each satisfying valuation of the encoding corresponds to a lasso-trace of $k$ states of the formula~\cite{Latvala+2004}. 
Then, we can use a propositional model counter~\cite{Bayardo+1997,Thurley+2006,Sharma+2019} to indirectly solve the LTL (bounded) model counting. 
%Notice that, 
The propositional encoding explodes exponentially as the bound is incremented, thus it can only be applied to small values of $k$. 
For the comparison, we additionally consider the approach that served us as motivation, presented in \cite{Degiovanni+2018}. 
It also attempts to approximate LTL model counting, by generating a regex from a formula, then fed to the string model counter ABC~\cite{Aydin+2015} to estimate the number of models.  
Encoding our reactive specifications into propositional constraints can quickly produce formulas beyond what exact model counters can analyse. 
Therefore, we generate ten sets, $S_0,\ldots, S_9$, with 50 random LTL formulas, feasible for all approaches, taken from a well-established benchmark~\cite{AaltaBenchmark}, used for assessing LTL SAT solvers \cite{Vardi+2015}. 
For each formula in each set, we compute the number of models, for $k \in [6..10]$, using the \emph{Exact} Model-Counting, the approach of \cite{Degiovanni+2018} (\textit{RE}), and our Approximate Model-Counting (\textit{ApMC}).
% We compute this for the bound $k$ from 6 to 10. 
Then, we rank the formulas in ascending order w.r.t. the number of models obtained with each technique. 
We compare these rankings to verify if the techniques preserve the ranking of the exact MC. 

\begin{table}[htp!]
\caption{Model Counting results.}
\vspace*{-0.5cm}
\begin{center} 
%\scalebox{.65}{
\resizebox{.34\textwidth}{!}{
\begin{tabular}{|c|r|r|r|r|r|}
\hline
    & Time & Time & Diff & Time & Diff \\
Set	& Exact  &  ApMC &  Exact - ApMC &  RE  &  Exact - RE\\
\hline
$S0$ & 3685 & 2 & 0  & 102  & 11\\
$S1$ & 8815 & 2 & 2  & 102  & 9\\
$S2$ & 5757 & 2 & 0  & 101  & 7\\
$S3$ & 5299 & 2 & 0  & 101  & 6\\
$S4$ & 5844 & 2 & 0  & 102  & 5 \\
$S5$ & 5665 & 2 & 0  & 102  & 6 \\
$S6$ & 5130 & 2 & 0  & 103  & 10\\
$S7$ & 5929 & 2 & 0  & 103  & 8\\
$S8$ & 8800 & 2 & 0  & 101  & 10\\
$S9$ & 6172 & 2 & 0  & 103  & 7\\
\hline
\end{tabular}
}
\end{center}
\label{tab:model-counting}
\end{table}
Table~\ref{tab:model-counting} shows the results of the comparison only for bound $k=10$ (other bounds are in the tool's site).
%For readability and space reasons, Table~\ref{tab:model-counting} shows the results of the comparison between the different counters; only contains the results for the highest bound, i.e., $k=10$ (other bounds are in the tool's site).
It reports the execution time in seconds and the difference between rankings provided by \emph{ApMC} and \emph{RE} w.r.t. the ranking of the \emph{Exact} model counter. 
%In that table, the fourth and sixth columns show the difference between rankings provided by \textit{Exact} and \textit{ApMC}, and \textit{Exact} and \textit{RE}, respectively. 
The results show that \textit{ApMC}, in 9/10 sets, produces the same ranking as \textit{Exact}.
In only 1 set it misclassified 2/50 formulas. 
Contrary, \textit{RE} misclassified 8 formulas per set on average. 
% The exact model counter required 470 seconds, for a bound of 6, to 6000 seconds, for a bound of 10, to analyze each set on average. 
\textit{Exact} required 470 seconds, for a bound of 6, to 6000 seconds, for a bound of 10, to analyze each set on average. 
% The approach of \cite{Degiovanni+2018} 
\textit{RE} required more than 100 seconds per set, while \textit{ApMC} required only 2 seconds per set. 
We assess scalability by testing approaches on 26 Table~\ref{tab:case-studies} specifications. \textit{Exact} quickly becomes infeasible, while \textit{RE} failed on 22/26 specifications, but \textit{ApMC} succeeded on all, even with large bounds.

%% file: related-work.tex
\section{Related Work}
\label{sec:related-work}

%Temporal Logics have been successfully used for specifying reactive systems \cite{MannaPnueli1995,Clarke2008}, and the problem of reactive synthesis from temporal logic specifications has been widely studied for many years. 
% The first proposals \cite{MannaWolper1984,EmersonClarke1982} rely on SAT solving for synthesising synchronisation skeletons for concurrent program given the specification in temporal logic. The work in \cite{PnueliRosner1989} is the first attempt in which the reactive synthesis problem is thought as a two players game, where the environment is treated as an adversary for the software component being synthesized. In \cite{PnueliRosner1989}, Pnueli and Rosner showed that the problem of reactive synthesis for general LTL specifications is double exponential in the length of the formula. Due to this intractable complexity in practice, various authors tried to define fragments of LTL in which the synthesis problem could be efficiently solved \cite{AlurLaTorre2001,Asarin+1998}. Piterman et al. \cite{Piterman+2006} presented an efficient symbolic synthesis algorithm for General Reactivity (1), a vey reach fragment of LTL that covers most of the commonly used LTL specifications patterns \cite{Dwyer+1999}. 
In reactive modeling, the system's
% behaviour is captured by state transition systems. At the same time, the 
expected properties are captured in temporal logics~\cite{MannaPnueli1992, MannaPnueli1995}, this is a typical setting in many essential activities, such as model checking \cite{Clarke2001}, property monitoring \cite{DBLP:journals/tosem/BauerLS11}, and model-based testing \cite{DBLP:journals/stvr/FraserWA09}. 
%The current system behavior must be fixed when it does not satisfy the specification. 
Detecting and finding flaws in specifications have been the focus on many studies~\cite{vanLamsweerdeLetier2000,Degiovanni+2016,Degiovanni+2018b}
Recent works present different techniques to automatically repair the system's specification
% the model that describes the system behaviour
~\cite{AlrajehCraven2014, Degiovanni+2014,Chatzieleftheriou+2015}. 
Unlike these approaches, which target the behavioural model, {\OurTool} aims to repair the declarative specification from which the synthesis tool will later generate an adequate model. 
LTL-Reactive synthesis has been studied for many years~\cite{MannaWolper1984,EmersonClarke1982,PnueliRosner1989,AlurLaTorre2001,Asarin+1998,Piterman+2006}. 
Many approaches have focused on diagnosing the cause of unsynthesisability by computing a core of assertions that makes the specification unreal~\cite{Schuppan2010} or by generating a counter-strategy showing how the environment can prevent the controller from satisfying the guarantees~\cite{RamanKress-Gazit2013}. 
Other approaches focus on undesirable properties of realisable specifications that affect the controllers' quality~\cite{DIppolito+2013}. 
{\OurTool} guarantees to produce satisfiable and realisable repairs, and the mentioned techniques can complement the analysis to assess and improve our repairs' quality. %get rid of this kind of undesired solutions.
%For instance, the works of  study the characterisation of anomalous controllers, that can force the environment to violate the assumptions. 

Recent approaches focus on inferring missing assumptions from unrealisable specifications~\cite{Chatterjee+2008, Alur+2013, CavezzaAlrajeh2016, Maoz+2019, li2011mining}. 
Their limitations are twofold: they work on LTL fragments, e.g., GR(1), and only attempt to solve unrealisability, adding assumptions, not considering that existing ones/guarantees might be incorrect, which is often the case~\cite{VanLamsweerde2009,vanLamsweerdeLetier2000, Alrajeh+2020}. 
We show in Section~\ref{sec:related-work-comparison} that {\OurTool} complements these techniques by being able to analyse general LTL specifications, changing both assumptions and guarantees, and providing more variants to repair unrealisability. 
%In contrast with previous approaches, we focus on solving the source of unrealisability, not only by adding new assumptions, but also by refining both assumptions and guarantees, from the specification.
% 
%They try to generate from the generated counter-strategies, a set of missing assumptions, such that a synthesisable specification is obtained. The work of \cite{Alur+2015} studied the problem of compositional reactive synthesis. They refine the specification of one component, by extracting information from the strategies and counter-strategies of other component's controller. 
 % ,Degiovanni+2016,Degiovanni+2018b}. 
%Many works from the RE community, e.g. \cite{vanLamsweerde+1998,vanLamsweerdeLetier2000,Degiovanni+2016,Degiovanni+2018b} have shown that the guarantees might be too ideal, partial and imprecise to start with, leaving out some exceptional conditions that may arise within its environment once implemented. 
%Recently, the work of \cite{Alrajeh+2020} states that the environment conditions are highly subject to change over time, that makes the assumptions to change, requiring later the adaptation of the entire goal model. \cite{Alrajeh+2020} presented a counterexample-guided learning procedure to generate an adapted goal model that satisfy general sanity properties of goal models, like correctness of the refinement between low-level and high-level goals. However, this work does not focus on the realizability property, that might be lost after changing the assumptions and guarantees from the specification. 
 % 
Program repair tools, like GenProg~\cite{LeGoues+2012} and DirectFix~\cite{MechtaevRoychoudhury2015}, use evolutionary algorithms to explore syntactical variants of the buggy program. %to be repaired. 
These algorithms also aim to guide the search toward repairs similar to the input program.  
{\OurTool} uses LTL model counting to measure the semantic distance of the candidates concerning the initial specification. 
We show that existing LTL model counting tools~\cite{Finkbeiner+2014, Degiovanni+2018} quickly reach their scalability limits. Thus, we developed an approach to approximate it, proving that it is more efficient than the mentioned methods. %and scaled well enough to analyse the specifications considered in our evaluation.

%Our work applies model counting to specifications formally captured as LTL formulas. 
%The work in \cite{Finkbeiner+2014} deals with the LTL model counting problem, but it is restricted to safety properties, making it unsuitable for our purposes. 
%The alternative of reducing LTL formulas to propositional formulas led us to constraints with thousands of variables that cannot be efficiently handled by efficient propositional model counters (e.g., \texttt{RelSAT}~\cite{Bayardo+1997}, \texttt{cachet}~\cite{Sang+2004} and \texttt{sharpSAT}~\cite{Thurley+2006}). 
%The work in \cite{Degiovanni+2018} takes an LTL formula and generates a regex that characterises prefixes of accepting lasso-traces of the formula, then fed to the string model counter ABC~\cite{Aydin+2015}. This regex generation involves intermediate steps (e.g., determinisation and minimisation), that are rather expensive. With a similar motivation, we developed an alternative approach to approximate the LTL model counting problem, that proved to be much more efficient than these related techniques.

%% file: conclusion.tex
%!TEX root=unreal-repair.tex
\section{Conclusion}
\label{sec:conclusion}
% The provision of automated mechanisms to assist engineers in identifying and resolving sources of unrealisability in temporal logic specifications has been the focus of many studies in the past years. 
% In this paper we presented an evolutionary approach that automatically searches for repairs of unrealisable specifications. 
% As opposed to previous methods that typically focus their analyses on the identification of missing assumptions, our approach aims at modifying both assumptions and guarantees. The key aim is to generate realisable versions that are as similar as possible to the given unrealisable ones. 
% We defined the notions of syntactic and semantic similarity, essentials in guiding the algorithm to quality solutions. 
% Our approach succeeded in repairing several case studies taken from literature and established benchmarks, and managed to generate solutions in line with manually and automated fixes. 
%There are a few lines of future work that we are interest in. In particular, we plan to investigate the possibility of using the information provided by the counter-strategies generated from unrealisable  candidate solutions, to improve the election of the mutation operator during the evolution of the genetic algorithm.

%Providing automated mechanisms to assist engineers in identifying and resolving sources of unrealisability in temporal logic specifications has been the focus of many studies in the past years. 
This paper presents {\OurTool}, a \textit{search-based} approach to repair unrealisable specifications. 
Compared to previous methods that typically focus their analyses on identifying missing assumptions, {\OurTool} aims to modify assumptions and guarantees. 
The key aim is to generate realisable versions close to the given unrealisable ones. 
We defined syntactic and semantic similarity notions, essentials in guiding the algorithm towards quality solutions. {\OurTool} succeeded in repairing several case studies from the literature, established benchmarks, and managed to generate solutions in line with manual and automated fixes.